\documentclass[showpacs,superscriptaddress,nofootinbib,floatfix,11pt]{revtex4-2}
\usepackage{setspace}
\usepackage[utf8]{inputenc}
\usepackage{float}
\usepackage{amsmath,amssymb,amsfonts,bm}
\usepackage{mathrsfs}
\usepackage{graphicx}
\usepackage{multirow}
\usepackage[usenames,dvipsnames]{color}
\allowdisplaybreaks
\usepackage[
colorlinks=true,
linkcolor=blue,
breaklinks=true,
citecolor=blue]{hyperref}

\usepackage{slashed}

\newcommand{\be}{\begin{equation}} 
\newcommand{\ee}{\end{equation}}
\newcommand{\bea}{\begin{eqnarray}} 
\newcommand{\eea}{\end{eqnarray}}
\newcommand{\beas}{\begin{eqnarray*}} 
\newcommand{\eeas}{\end{eqnarray*}}
\newcommand{\al}{&}
\newcommand{\lag}{\mathcal{L}}

\newcommand{\nno}{\nonumber\\}

\newcommand{\w}{\omega}

\newcommand{\mb}{m_{\Lambda_b}}
\newcommand{\mcg}{m_{\Lambda_{c,1/2}^*}}
\newcommand{\mce}{m_{\Lambda_{c,3/2}^*}}

\newcommand{\hh}{\widetilde{h}}

\newcommand{\lamcg}{\Lambda_c(2595)^+}
\newcommand{\lamce}{\Lambda_c(2625)^+}

\newcommand{\lcg}{\Lambda_{c,1/2^-}^*}
\newcommand{\lce}{\Lambda_{c,3/2^-}^*}
\newcommand{\lambb}{\Lambda_b}

\newcommand{\omc}{\mathcal{O}(\Lambda_\text{QCD}/m_c)}
\newcommand{\fzero}{f_0^{(\frac12^-)}}
\newcommand{\gzero}{g_0^{(\frac12^-)}}
\newcommand{\fplus}{f_+^{(\frac12^-)}}
\newcommand{\gplus}{g_+^{(\frac12^-)}}
\newcommand{\fperp}{f_\perp^{(\frac12^-)}}
\newcommand{\gperp}{g_\perp^{(\frac12^-)}}

\newcommand{\ific}{\affiliation{Instituto de F\'isica Corpuscular (centro mixto CSIC-UV), \\
Institutos de Investigaci\'on de Paterna, Apartado 22085, 46071, Valencia, Spain}}

\newcommand{\usal}{\affiliation{Departamento de F\'isica Fundamental e IUFFyM, Universidad de Salamanca, \\
Plaza de la Merced s/n, E-37008 Salamanca, Spain}}

\begin{document}

\title{Is the $\Lambda_c(2625)^+$ the heavy quark spin symmetry partner of the $\Lambda_c(2595)^+$?}

\begin{abstract}
We use a $\omc$ heavy quark effective theory scheme, where only $\mathcal{O}(\Lambda_\text{QCD}/m_b)$ and perturbative QCD short distance corrections are neglected, to study the  matrix elements of the scalar, pseudoscalar, vector, axial-vector and tensor currents between the $\lambb$ ground state and the odd parity charm  $\lamcg$ and $\lamce$ resonances. 
  We show that in the near-zero recoil regime,  the scheme describes reasonably well, taking into account uncertainties, the results for the 24  form-factors obtained in lattice QCD (LQCD) just in terms of only 4 Isgur-Wise (IW) functions.  We also find  some support for  the possibility that the $\lamcg$ and $\lamce$ resonances  might form a  heavy-quark spin symmetry (HQSS) doublet. However, we argue that the available LQCD description of these two resonances is not accurate enough to disentangle the possible effects of the $\Sigma_c \pi$ and $\Sigma_c^*\pi$ thresholds, located only a few MeV above their position,  and that it cannot be ruled out that these states are not HQSS partners. Finally, we study the ratio $\frac{d\Gamma[\Lambda_b\to \lcg \ell \bar\nu_\ell]/dq^2}{d\Gamma[\Lambda_b\to \lce\ell \bar\nu_\ell]/dq^2}$ of the Standard Model differential semileptonic decay widths, with $q^\mu$ the momentum transferred between the initial and final hadrons. We provide a natural explanation  for  the existence of large deviations,  near the zero recoil, of this  ratio from 1/2 (value predicted in the infinite heavy quark mass limit, assuming  that the $\lcg$ and $\lce$ are the two members of a HQSS doublet) based on  S-wave contributions to the $\lambb\to \lcg$ decay amplitude driven by a sub-leading IW function.
\end{abstract}

\author{Meng-Lin Du}\email{du.menglin@ific.uv.es}
\ific

\author{Eliecer Hern\'andez}\email{gajatee@usal.es}
\usal

\author{Juan Nieves}\email{jmnieves@ific.uv.es}
\ific

\maketitle

\section{Introduction}

Heavy quark symmetry plays an important role in our understanding of low-energy strong interactions and the classification of the  heavy-light hadronic spectrum. In the infinite heavy quark mass limit ($m_Q\to \infty$), the degrees of freedom of the infinitely massive heavy quark decouple from the light quark ones, and hence the heavy and light degrees of freedom are separately conserved. The dynamics of hadrons containing a heavy quark is blind to the flavour and  spin of the heavy quark, with the latter exhibiting an SU(2)$_Q$ pattern known as the heavy-quark spin symmetry. Light degrees of freedom (ldof) with  spin-parity $j_{\rm ldof}^P$ yield to a degenerate doublet under rotations of the heavy quark spin $s_Q$. The quantum numbers of this doublet are $J^P=\left[j_{\rm ldof}\pm \frac12\right]^P$ since the spin-parity of the heavy quark is $\frac{1}{2}^+$. Note here that the ldof quantum numbers contain both the orbital angular-momentum  and spin parts. 

In the real world, HQSS is broken due to the large ($m_Q\gg \Lambda_\text{QCD}$), but finite mass of the  heavy quark, e.g. $c$ and $b$ quarks. These doublets are not exactly degenerate with a hyperfine-splitting of order $\Lambda_\text{QCD}/m_Q$. The similar  masses of the isoscalar odd parity resonances $\lamcg$ and $\lamce$, with $J^P=\frac{1}{2}^-$ and $\frac{3}{2}^-$ respectively, makes them promising  candidates for the lightest charmed baryon doublet with $j_{\rm ldof}^P=1^-$. This assignment  has been widely adopted in the literature, see e.g. Refs.~\cite{Leibovich:1997az, Pervin:2005ve, Yoshida:2015tia, Nagahiro:2016nsx, Boer:2018vpx, Meinel:2021rbm, Papucci:2021pmj}. Various constituent quark models (CQMs) predict a nearly degenerate pair of P-wave charmed isoscalar baryons $\Lambda_c^*$ with $J^P=\frac{1}{2}^-$ and $\frac32^-$, respectively, which masses are close to those of the $\lamcg$ and $\lamce$ \cite{Migura:2006ep,Garcilazo:2007eh,Roberts:2007ni,Yoshida:2015tia}. Two different orbital excitations are considered in Ref.~\cite{Yoshida:2015tia}, driven by the so-called  $\lambda$- and $\rho$-degrees of freedom. While the former accounts for the excitation between the heavy quark and the  light quark subsystem, the latter considers the excitation between the two light quarks. Owing to the big difference between the heavy- and light-quarks masses, the low-lying states are dominated by the lowest $\lambda$-mode excitation and the mixture with $\rho$ excitations is small \cite{Yoshida:2015tia}. In this picture, the $\lamcg$ and $\lamce$ correspond to the HQSS doublet associated to ($\ell_\lambda=1$, $\ell_\rho=0$) with  total ldof spin $s_{\rm ldof}=0$, leading to $j_{\rm ldof}^P=1^-$. The predicted decay widths of these resonances within this scheme are found to be consistent with  data \cite{Nagahiro:2016nsx,Arifi:2017sac}.

The $\lamcg$ and $\lamce$ have also been  described in hadronic-molecular models, as the counterparts of the $\Lambda(1405)$ and $\Lambda(1520)$ with the strange quark replaced by the charm one  \cite{Lutz:2003jw,Tolos:2004yg,Hofmann:2005sw,Mizutani:2006vq,Hofmann:2006qx,Garcia-Recio:2008rjt,Romanets:2012hm,Lu:2014ina,Lu:2016gev}. Since the $\Sigma_c$ and $\Sigma_c^*$ form the ground $j_{\rm ldof}^P=1^+$ HQSS doublet, the quantum numbers of the ldof of the S-wave $\Sigma_c\pi$ and $\Sigma_c^*\pi$ pairs are $1^-$ for both cases. In Ref.~\cite{Lutz:2003jw}, the $\lamcg$ is dynamically generated from the S-wave scattering of the Goldstone bosons off the $J^P=\frac12^+$ charmed baryon octet. A similar treatment is performed in Ref.~\cite{Lu:2014ina}, where the masses of the $\lamcg$ and $\lamce$ are reproduced as $\Sigma_c\pi$ and $\Sigma_c^*\pi$ bound states by fine-tuned parameters. 
A coupled-channel approach including the $DN$ is studied in Ref.~\cite{Tolos:2004yg} for the $\lamcg$, and it is extended to the $\lamce$ sector subsequently in Refs.~\cite{Hofmann:2005sw,Mizutani:2006vq}. HQSS, however, is not respected since the $D^*N$ channel is not considered and   the $D$ and $D^*$ mesons form the ground $j_{\rm ldof}^P=1/2^-$ HQSS doublet in the meson sector. The first molecular description of the $\lamcg$ and $\lamce$ respecting HQSS was provided in Refs.~\cite{Garcia-Recio:2008rjt,Romanets:2012hm}, where the SU(3) Weinberg-Tomozawa chiral Lagrangian is extended to SU(6)$_{\rm lsf}\times$SU(2)$_Q$, with lsf standing here for the spin-flavor symmetry in the light sector. The model is supplemented by a pattern of symmetry breaking corrections and using the particular renormalization scheme proposed in Refs.~\cite{Hofmann:2005sw,Hofmann:2006qx}. One $J^P=\frac32^-$ state is dynamically generated by the $\Sigma_c^*\pi$-$D^*N$ coupled-channel dynamics, which is identified in  \cite{Garcia-Recio:2008rjt} with the $\lamce$, even though its mass is about 40 MeV larger and it is significantly wider than the physical resonance. Thus within this scheme, the $\lamce$ would be the counterpart of the $\Lambda(1520)$ in the charm sector. Interestingly, the model of Refs.~\cite{Garcia-Recio:2008rjt,Romanets:2012hm} produce two $J^P=\frac12^-$ states generated near the nominal position of the $\lamcg$. One is narrow and it strongly couples to $DN$ and especially to $D^*N$, with a small mixing with  $\Sigma_c\pi$. It is identified with the  $\lamcg$ resonance in Refs.~\cite{Garcia-Recio:2008rjt,Romanets:2012hm}. The other $J^P=\frac12^-$  molecular state, however, is quite broad because of the sizable coupling to the open channel $\Sigma_c\pi$. The two states would be  analogous to those forming the two-pole structure of the $\Lambda(1405)$ \cite{Oller:2000fj,Garcia-Recio:2002yxy, Garcia-Recio:2003ejq, Hyodo:2011ur}, where the states couple to $\Sigma\pi$ and $\bar KN$, respectively \cite{Hyodo:2011ur}.  A big difference, between the charm and strange sectors,  is that while $D^*N$ plays a crucial role in the charm sector within the scheme of Refs.~\cite{Garcia-Recio:2008rjt,Romanets:2012hm}, the $\bar K^*N$ channel is not considered in the chiral unitary approaches~\cite{Oller:2000fj,Garcia-Recio:2002yxy, Garcia-Recio:2003ejq, Hyodo:2011ur} because of the large $\bar K^*$-$\bar K$ mass gap. 

In Ref.~\cite{Nieves:2019kdh}, it is stressed  that the narrow $J^P=\frac12^-$ $\lamcg$ found in ~\cite{Garcia-Recio:2008rjt,Romanets:2012hm} is mostly generated from the $DN$-$D^*N$ coupled-channel dynamics with a dominant $j_{\rm ldof}^P=0^-$ configuration. The small coupling of this state to the $\Sigma_c\pi$ channel is then a consequence of HQSS due to its small $j_{\rm ldof}=1^-$ component. The coupling of the broad $J^P=\frac12^-$ state to the $\Sigma_c\pi$, however, is larger than those to $DN$ and $D^*N$ and thus dominated by the $j_{\rm ldof}^P=1^-$ configuration. It means that the isoscalar $J^P=\frac32^-$ $\lamce$ state found in Refs.~\cite{Garcia-Recio:2008rjt,Romanets:2012hm} would be the HQSS partner of the broad $J^P=\frac12^-$ state with $j_{\rm ldof}^P=1^-$ \cite{Nieves:2019kdh}, instead of the observed $\lamcg$. A similar two-pole structure for the $J^P=\frac12^-$ sector is found in Ref.~\cite{Liang:2014kra} by making use of a SU(4) flavor extension of the local hidden gauge formalism.  In that work, an additional broad state around 2675 MeV is found in the isoscalar $J^P=\frac32^-$ sector with the single-channel $\Sigma_c^*\pi$ Weinberg-Tomozawa interaction, which would not be related to the $\lamce$.

The interplay between the $\Sigma_c^{(*)}\pi$-$D^{(*)}N$ baryon-meson pairs and the bare P-wave CQM states has been recently studied,  in the framework of an effective field theory, respecting heavy quark spin and chiral symmetries~\cite{Nieves:2019nol}. It is shown that the $\lamce$ should be viewed mostly as a dressed three-quark state, that is originated from a bare CQM state. The $\lamcg$, however, should be either dynamically generated by the chiral $\Sigma_c\pi$ interaction, or  the result of the $DN$-$D^*N$ coupled-channel dynamics with a $j_{\rm ldof}^P=0^-$ configuration, depending on the employed renormalization procedure. In any case, these two resonances would not be HQSS partners \cite{Nieves:2019nol}. This is because the bare $J^P=\frac32^-$ CQM state and the $\Sigma_c\pi$ threshold are located extremely close to the $\lamce$ and $\lamcg$, respectively, and thus play different roles in each sector.

HQSS also puts constraints on the form factors of currents containing heavy quarks, based on the observation that the current $\bar{Q}\Gamma q$ ($\bar{Q}\Gamma Q^\prime$) transforms as a spinor under SU(2)$_Q$ (SU(2)$_{Q^\prime}$ as well) \cite{Mannel:1990vg}. In particular in the heavy quark limit, the semileptonic decay of a $\lambb$ into the lowest HQSS-doublet with $j_{\rm ldof}^P=1^-$ can be described by a universal leading order Isgur-Wise (IW) function ($\sigma$) \cite{Isgur:1990pm}. Besides, at  zero recoil, the  weak-current matrix elements between a $\lambb$ and any  excited charmed baryon vanish \cite{Leibovich:1997az}. 
The form factors for the exclusive semileptonic $\lambb$ decays to the excited $\Lambda_c$ baryons were first obtained to  order $\omc$ in heavy quark effective theory (HQET) in Ref.~\cite{Roberts:1992xm}, and were improved to order $\mathcal{O}(\Lambda_\text{QCD}/m_b)$ in Refs.~\cite{Leibovich:1997az,Boer:2018vpx}, where the $\lamcg$ and $\lamce$ were regarded as the lowest-lying $j_{\rm ldof}^P=1^-$ doublet. At  order $\mathcal{O}(\Lambda_\text{QCD}/m_Q)$ (here $Q=c$ and $b$), there appear five  additional independent functions ($\sigma_1^{(c)}$, $\phi_\text{kin}^{(c)}$, $\phi_\text{kin}^{(b)}$, $\phi_\text{mag}^{(c)}$ and $\phi_\text{mag}^{(b)}$) and two low energy constants (LECs) $\bar \Lambda$ and ${\bar \Lambda}^\prime$, following the notation introduced in Ref.~\cite{Leibovich:1997az}. The LECs  (${\bar \Lambda}^{(\prime)}$) denote the energy in the hadron of the light degrees of freedom  in the $m_Q\to \infty$ limit. The functions $\phi_\text{kin}^{(c)}$ and $\phi_\text{kin}^{(b)}$ can be re-absorbed, neglecting $\mathcal{O}(\Lambda_\text{QCD}^2/m_Q^2)$, in the leading order IW function $\sigma$,  common for both decays. Therefore,  to describe the various form factors of the currents between the $\Lambda_b$ and the $\lcg$ and $\lce$ states at  order $\mathcal{O}(\Lambda_\text{QCD}/m_Q)$, only three of these sub-leading functions and the two LECs $\bar \Lambda$ and ${\bar \Lambda}^\prime$ are needed~\cite{Roberts:1992xm,Nieves:2019kdh}. Furthermore, if the $\mathcal{O}(\Lambda_\text{QCD}/m_b)$ corrections are neglected, and one only considers up to  order $\omc$, all the form factors for the $\Lambda_b \to \lcg, \lce$ transitions  
are given in terms of only three independent functions, $\sigma(\w), \phi_\text{kin}^{(c)}$ and $\sigma_1^{(c)}$,  and the $\bar \Lambda$ and ${\bar \Lambda}^\prime$ LECs (see for instance also Eqs.~(31)-(33) of Ref.~\cite{Nieves:2019kdh}). The $\Lambda_\text{QCD}/m_b$ or $\Lambda_\text{QCD}^2/m_c^2$ contributions, not taken into account in this limit, are expected to be
smaller than the theoretical uncertainties induced by the errors on $({\bar\Lambda}-{\bar\Lambda}^\prime)$.

Moreover, the semileptonic form factors between the $\lambb$ and a final $J^P=\frac12^-$ charm-baryon, but with $j_{\rm ldof}^P=0^-$, denoted as $\Lambda_{c,1/2^-}^{\prime *}$, vanish in the $m_Q\to\infty$ limit. The unique non-vanishing correction at  order $\mathcal{O}(\Lambda_\text{QCD}/m_c)$ comes from the chromomagnetic operator, which can be described by a universal function with a structure of the type $\epsilon_{\mu\nu\rho\tau}\sigma^{\mu\nu}v^\rho v^{\prime\tau}$ with $v$ and $v^\prime$ the four-velocities of the $\lambb$ and $\Lambda_{c,1/2^-}^{\prime *}$, respectively~\cite{Leibovich:1997az}. Thus, the different HQSS  pattern of form factors for the $\lambb$ semileptonic decay into a charm $j_{\rm ldof}^P=1^-$  doublet or a $j_{\rm ldof}^P=0^-$  singlet might be used to test whether the ldof configuration in the $\lamcg$  corresponds mainly to any of these quantum numbers, provided that the decays are measured.

Given the lack of data for the form-factors of the $\Lambda_b\to\lamcg$ and $\lambb\to\lamce$  semileptonic decays, the lattice QCD (LQCD) simulations carried out in Refs.~\cite{Meinel:2021rbm,Meinel:2021mdj} provide valuable information  from  first principles\footnote{The form factors for the $\lambb\to\lcg$ and $\lambb\to\lce$ transitions have  also been  investigated within the framework of  constituent quark-models, see e.g., in Refs.~\cite{Pervin:2005ve,Gutsche:2017wag,Becirevic:2020nmb}.}. (In the  second of the  references, the exact zero-recoil rotational symmetry relations among the different form factors are imposed to ensure the correct behavior of the angular observables). Using LQCD form factors~\cite{Bowler:1997ej,Gottlieb:2003yb,Detmold:2015aaa,Datta:2017aue}, the line shape of the $\lambb\to\Lambda_c \mu^-\bar{\nu}_\mu$ decay, with $\Lambda_c$ the $J^P=1/2^+$ ground state charm baryon for which $j_{\rm ldof}^P=0^+$,    has been found to be consistent with the LHCb measurement \cite{LHCb:2017vhq}.  This gives strong support to the LQCD calculation of semileptonic form factors. Note, however, that the form factors for the $\lambb\to \Lambda_c$ and those for the $\lambb\to \lcg, \lce$ transitions are not related at all, since the ldof in the final charm baryon have different configurations. In Ref.~\cite{Leibovich:1997az}, branching fractions and heavy quark sum rules for $\lambb$ decays to $\lamcg$ and $\lamce$ (identified as the HQSS-doublet of $j_{\rm ldof}^P=1^-$) within the Standard Model (SM) were evaluated in the large $N_c$ limit of QCD, using the bound state soliton picture.

The relevant expressions for the matrix elements, up to order $\alpha_s$ and $\Lambda_\text{QCD}/m_Q$ in HQET, to test  lepton flavour universality are provided in Ref.~\cite{Boer:2018vpx}. However in that work, only rough estimates of the form-factors, obtained from the zero recoil sum
rule~\cite{Shifman:1994jh,Bigi:1994ga}, could be used. New physics signatures for various  $b\to c \ell \bar{\nu}_\ell $ four Fermi interactions in the baryon sector are also investigated in Ref.~\cite{Papucci:2021pmj}, using a HQET-based parametrization of the
form factors to  existing quark model results~\cite{Pervin:2005ve}. As previously
noted also in Ref.~\cite{Meinel:2021rbm}, the results of Ref.~\cite{Papucci:2021pmj} show a tension between LQCD data and HQET predictions to order $\alpha_s$ and $\Lambda_\text{QCD}/m_Q$, and points out to the existence of  unexpectedly large HQET-violating terms, potentially large  $1/m_c^2$
corrections near zero recoil, in the LQCD form-factors.  Further studies of this issue
are thus needed, since it is still an open question how well  HQSS works for  the semileptonic $\Lambda_b\to\lamcg$ and $\lambb\to\lamce$ transitions, and if the $\lamcg$ and $\lamce$ form a HQSS doublet~\cite{Nieves:2019nol}.

In this work, we will try to answer some of these questions using the LQCD  form factors recently obtained in Refs.~\cite{Meinel:2021rbm,Meinel:2021mdj}. In Sec.~\ref{sec:form}, we will provide the explicit definition of the $\lambb\to\Lambda_c^*$ semileptonic form factors used in this work, and the relations between them and those obtained in  Refs.~\cite{Meinel:2021rbm,Meinel:2021mdj}. We briefly review  the form factors in HQET, up to order $\omc$, in  Sec.~\ref{sec:hqss}. We test how well HQSS describes these form factors and perform a detailed numerical analysis in Sec.~\ref{sec:num}. Section~\ref{sec:summary} includes a brief summary of the main results of this work. Finally in the Appendices~\ref{app:LQCDFF} and \ref{app:ff:hqss}, we provide  relations between the different sets of form-factors considered in this work.

\section{Form factors for $\lambb\to\Lambda_c^*$ semileptonic transitions}\label{sec:form}

The form factors for the $\lambb\to\lamcg \ell \bar\nu_\ell$ [$\lambb\to\lamce \ell \bar\nu_\ell$]  semileptonic transitions parameterize  the matrix elements of the local currents $[\bar{c}(0)\Gamma b(0)]$ between the $\lambb$ and the $\lamcg$ [$\lamce$] states. Following the notation of Ref.~\cite{Leibovich:1997az} for the vector and axial-vector currents and extending it to the scalar, pseudoscalar, and tensor currents, we have for the $\lambb\to\lamcg$
\bea\label{eq:formfactor12}
\langle \lamcg (p^\prime,s^\prime)| \bar{c}b|\Lambda_b (p,s) \rangle \al = \al d_S \bar{u}_c(p^\prime,s^\prime) \gamma_5 u_b(p,s), \nno
\langle \lamcg (p^\prime,s^\prime)| \bar{c}\gamma_5 b|\Lambda_b(p,s) \rangle \al = \al d_P \bar{u}_c(p^\prime,s^\prime)  u_b(p,s), \nno
\langle \lamcg (p^\prime,s^\prime)| \bar{c}\gamma^\mu b|\Lambda_b (p,s) \rangle \al = \al \bar{u}_c(p^\prime,s^\prime)\left[ d_{V_1} \gamma^\mu + d_{V_2}\frac{p^\mu }{\mb} + d_{V_3} \frac{p^{\prime\mu} }{\mcg} \right] \gamma_5 u_b(p,s), \nno
\langle \lamcg (p^\prime,s^\prime)| \bar{c}\gamma^\mu \gamma_5 b|\Lambda_b (p,s) \rangle \al = \al \bar{u}_c(p^\prime,s^\prime)\left[ d_{A_1} \gamma^\mu + d_{A_2} \frac{p^\mu }{\mb} + d_{A_3}\frac{p^{\prime\mu}}{\mcg}\right] u_b(p,s), \nno 
\langle \lamcg (p^\prime,s^\prime)| \bar{c}\sigma^{\mu\nu} \gamma_5 b|\Lambda_b (p,s) \rangle \al = \al\bar{u}_c(p^\prime,s^\prime)\Big[ i\frac{d_{T_1}}{\mb^2}(p^\mu p^{\prime\nu}-p^\nu p^{\prime\mu}) + i \frac{d_{T_2}}{\mb}(\gamma^\mu p^\nu - \gamma^\nu p ^\mu ) \nno 
\al   \al + i\frac{d_{T_3}}{\mb}(\gamma^\mu p^{\prime \nu} - \gamma^\nu p^{\prime\mu}) + d_{T_4}\sigma^{\mu\nu} \Big] u_b(p,s),
\eea
where $\mcg$ is the mass of $\lamcg$, and $u_b(p,s)$ and ${u}_c(p^\prime,s^\prime)$ are the spinors of $\lambb$ and $\lamcg$ baryons, respectively ($p$ and $p^\prime$ are four-momenta, while $s$ and $s^\prime$ are spin indices). The form-factors $d_{S,P,V_i,A_i,T_i}$ are scalar functions of  $q^2=(p-p')^2$, or equivalently $  \omega =(\mb^2+\mcg^2-q^2)/(2\mb\mcg)$. 

The form factor decomposition for the $\langle \lamcg (p^\prime,s^\prime)| \bar{c}\sigma^{\mu\nu}  b|\Lambda_b (p,s) \rangle $ matrix element can be straightforwardly obtained from that of the $\sigma^{\mu\nu}\gamma_5$ operator by making use of $\sigma^{\mu\nu}\gamma_5 = -\frac{i}{2}\epsilon^{\mu\nu\lambda\rho}\sigma_{\lambda\rho}$, with the convention $\epsilon_{0123}=+1$. 

Likewise, the $\lambb\to\lamce$ matrix elements can be parameterized as
\bea\label{eq:formfactor32}
\langle \lamce(p^\prime,s^\prime)|\bar{c}b| \lambb(p,s)\rangle \al = \al l_S \frac{p^\lambda}{\mb} \bar{u}_{c,\lambda}(p^\prime,s^\prime) u_b(p,s), \nno
\langle \lamce(p^\prime,s^\prime)|\bar{c}\gamma_5 b| \lambb(p,s)\rangle \al = \al l_P \frac{p^\lambda}{\mb} \bar{u}_{c,\lambda}(p^\prime,s^\prime) \gamma_5 u_b(p,s), \nno 
\langle \lamce(p^\prime,s^\prime)|\bar{c}\gamma^\mu b|\lambb(p,s)\rangle \al = \al \bar{u}_{c,\lambda}(p^\prime,s^\prime) \Bigg[ \frac{p^\lambda}{\mb}\left( l_{V_1}\gamma^\mu + l_{V_2}\frac{p^\mu }{\mb} + l_{V_3}\frac{p^{\prime\mu}}{\mce}\right) \nno
\al  \al +  l_{V_4} g^{\lambda\mu} \Bigg] u_b(p,s), \nno 
\langle \lamce(p^\prime,s^\prime)|\bar{c}\gamma^\mu\gamma_5 b|\lambb(p,s)\rangle \al = \al \bar{u}_{c,\lambda}(p^\prime,s^\prime) \Bigg[ \frac{p^\lambda}{\mb}\left( l_{A_1}\gamma^\mu + l_{A_2}\frac{p^\mu }{\mb} + l_{A_3}\frac{p^{\prime\mu}}{\mce}\right) 
\nno
\al \al + l_{A_4} g^{\lambda\mu} \Bigg] \gamma_5 u_b(p,s), \nno 
\langle \lamce(p^\prime,s^\prime)|\bar{c}\sigma^{\mu\nu} b|\lambb(p,s)\rangle \al = \al \bar{u}_{c,\lambda}(p^\prime,s^\prime) \bigg[ \frac{p^\lambda}{\mb}\Big(  i\frac{l_{T_1}}{\mb^2}(p^\mu p^{\prime\nu}-p^\nu p^{\prime\mu}) + i\frac{l_{T_2}}{\mb}(\gamma^\mu p^\nu -\gamma^\nu p^\mu)  \nno 
\al \al + i \frac{l_{T_3}}{\mb}(\gamma^\mu p^{\prime\nu}-\gamma^\nu p^{\prime\mu}) + l_{T_4}\sigma^{\mu\nu}\Big) + i l_{T_5}(g^{\lambda\mu}\gamma^\nu-g^{\lambda\nu}\gamma^\mu ) \nno 
\al \al + i \frac{l_{T_6}}{\mb}(g^{\lambda\mu}p^\nu - g^{\lambda\nu} p^\mu ) \bigg]u_b(p,s),
\eea
where $\mce$ is now  the mass of $\lamce$, $u_{c,\lambda}(p^\prime,s^\prime) $ is the Rarita-Schwinger spinor for the spin-$\frac32$ $\lamce$, which satisfies $p^{\prime\lambda}u_{c,\lambda}=\gamma^\lambda u_{c,\lambda}=0$, and the form-factors $l_{S,P,V_i,A_i,T_i}$ are scalar functions of  $q^2=(p-p')^2$, or $\omega$, which is now expressed in terms of $\mce$.   Note there is also one more independent structure $i\frac{l_{T_7}}{\mb}(g^{\lambda\mu}p^{\prime\nu}-g^{\lambda\nu}p^{\prime\mu})$ in the  form factor decomposition of the tensor operator, which can not be eliminated simply by the equation of motion or transversality conditions. However, it is shown in Ref.~\cite{Papucci:2021pmj} that the combination of operators 
\be
\mathcal{K}_{\mu\nu} = \bar{u}_{c}^{\lambda} \Big[ v_\lambda \Big( \sigma_{\mu\nu} - i  (v^{\prime}_{\mu}\gamma_\nu -v^{\prime}_{\nu}\gamma_\mu) \Big) + i \Big(g_{\lambda\mu}\big[(\w+1)\gamma_\nu-{v_\nu}-v^{\prime}_{\nu}\big] - g_{\lambda\nu}\big[(\w+1)\gamma_\mu-{v_\mu}-v^{\prime}_{\mu} \big]\Big) \Big] u_b
\ee
does not contribute to physical amplitudes. Here we have introduced the notations $v_\mu = p_\mu/\mb$, $v_\mu^\prime=p_\mu^\prime/\mce$ and $\w=v\cdot v^\prime$, determined by $q^2$. Thus, we can set to zero the contribution of $l_{T_7}$ to the physical amplitudes by redefining $l_{T_i}$ for $i=3,4,5, 6$. 

On the other hand, as in the case of the $\lamcg$, the   $\langle \lamce(p^\prime,s^\prime)|\bar{c}\sigma^{\mu\nu}\gamma_5 b|\lambb(p,s)\rangle $ form-factors can be obtained from the tensor ones introduced in the decomposition of the matrix element of the $\sigma^{\mu\nu}$ operator.

The helicity form factors for $\lambb\to \lamcg$ and $\lambb\to\lamce$ determined in the LQCD simulation carried out in Refs.~\cite{Meinel:2021mdj,Meinel:2021rbm} are linear combinations of those introduced above, and the relation between both sets
of form-factors is given in Appendix~\ref{app:LQCDFF}.

\section{$\lambb\to\Lambda_c^*$ form factors  and heavy-quark effective theory}\label{sec:hqss}

In this section, we will provide the form factors for the semileptonic decay of the $\lambb$ into the members of the  $j_{\rm ldof}^P=1^-$ HQSS doublet $\lcg$ and $\lce$ in HQET up to order $\omc$ and neglecting QCD short range
logarithms~\cite{Papucci:2021pmj}.

In the infinite heavy quark limit ($m_Q\to\infty$), the form factors $d_i$ and $l_i$ in Eqs.~\eqref{eq:formfactor12} and \eqref{eq:formfactor32} for the $\lambb$ decay into the two members ($\Lambda_c^*$) of the $j_{\rm ldof}^P=1^-$ HQSS-doublet can be described by a universal IW function~\cite{Isgur:1990pm}. In this limit, the ground state $\lambb$ is a HQSS-singlet with $j_{\rm ldof}^P=0^+$, and thus can be described by a Dirac spinor $u_b(v)$, with $v$ the velocity of the $\lambb$, satisfying the condition $\slashed{v}u_b(v)=u_b(v)$~\cite{Falk:1991nq}. For the $j_{\rm ldof}^P=1^-$ doublet with velocity $v^\prime$, the ldof are represented by a vector $A^\mu$ subject to the transversality condition $v^\prime\cdot A=0$. Then, the $\Lambda_c^*$ doublet can be introduced by the multiplet spinor 
\bea
\mathcal{U}^\mu_c(v^\prime)= A^\mu u_h(v'),
\eea
where the spinor $u_h$ describes the heavy  quark $c$ obeying $\slashed{v}^\prime u_h=u_h$. Note that  $\mathcal{U}^\mu_c(v^\prime)$ is not an irreducible representation under the Lorentz group, instead it contains both spin $3/2=(1+1/2)$ and $1/2=(1-1/2)$ components,
\bea\label{eq:spinors}
\mathcal{U}^\mu_c(v^\prime)= u_c^{\mu}(v^\prime) + \frac{1}{\sqrt{3}}(\gamma^\mu+v^{\prime\mu})\gamma_5 u_c(v^\prime),
\eea
where $u_c^{\mu}(v^\prime)$ and $u_c(v^\prime)$ are the Rarita-Schwinger and Dirac spinors for $J^P=\frac32^-$ and $\frac12^-$ states, respectively. Note that $v^\prime\cdot\mathcal{U}_c(v^\prime)=0$, $\slashed{v}^\prime\mathcal{U}_c(v^\prime)=\mathcal{U}_c(v^\prime)$, and $\gamma_\mu u_c^\mu(v^\prime)=0$. Then one can easily obtain that in the $m_Q\to\infty$ limit, the most general
form for the matrix element respecting HQSS is (i.e., being invariant under arbitrary separate rotations of the $b$- and $c$-quarks spins~\cite{Nieves:2019kdh})
\bea\label{eq:ff:lo}
\langle \Lambda_{c}^*;j_{\rm ldof}^P=1^-|\bar{h}_{v^\prime}^{(c)}\Gamma h_v^{(b)}|\Lambda_b\rangle = \sigma(\w) v_\lambda \bar{\mathcal{U}}_c^\lambda (v^\prime) \Gamma u_b(v).
\eea
where $h_v^{(Q)}$ is the heavy quark field in HQET, $\Gamma$ is a Dirac matrix, and $\sigma(\w)$ is the dimensionless leading IW function. Using Eqs.~\eqref{eq:spinors} and \eqref{eq:ff:lo}, one finds
\bea\label{eq:lo2}
\langle \lcg |\bar{h}_{v^\prime}^{(c)}\Gamma h_v^{(b)}|\lambb\rangle \al = \al \frac{1}{\sqrt{3}}\sigma(\w)\bar{u}_c(\slashed{v}-\w)\gamma_5\Gamma u_b, \nno
\langle \lce |\bar{h}_{v^\prime}^{(c)}\Gamma h_v^{(b)}|\lambb\rangle \al = \al \sigma(\w) v_\lambda \bar{u}_c^\lambda \Gamma u_b.
\eea
and thus, it is straightforward to obtain~\cite{Nieves:2019kdh} the $d_i$ and $l_i$ form-factors  in terms of $\sigma(\w)$. 

At order $\mathcal{O}(\Lambda_\text{QCD}/m_Q)$ , there are corrections originating from the matching of the $b\to c$ flavor changing current onto the effective theory (HQET) and from order $\mathcal{O}(\Lambda_\text{QCD}/m_Q)$ effective Lagrangian~\cite{Leibovich:1997az}. Considering for simplicity only $\omc$ contributions, this is to say keeping invariance under rotations of the spin of the  $b$ quark, we have first for the current corrections~\cite{Neubert:1993mb} 
\bea\label{eq:currentO}
\bar{c}\Gamma b = \bar{h}_{v^\prime}^{(c)} \left( \Gamma -\frac{i}{2m_c} \overleftarrow{\slashed{D}} \Gamma\right) h_v^{(b)},
\eea
where $D$ is the gauge covariant derivative. The charm quark next-leading order effective Lagrangian contains the  kinetic energy and the chromomagnetic terms~\cite{Neubert:1993mb} 
\bea\label{eq:lagO}
\delta \lag_v^{\prime(c)} = \frac{1}{2m_c}\left( \bar{h}_{v^\prime}^{(c)}(iD_\perp)^2 + \frac{g_s}{2}\sigma\cdot G\right) h_{v^\prime}^{(c)}.
\eea
The operator that appears in the correction of Eq.~\eqref{eq:currentO} can be parameterized as~\cite{Roberts:1992xm} 
\newcommand{\vp}{v^\prime}
\bea\label{eq:Dterm}
\bar{h}_{\vp}^{(c)}i\overleftarrow{D}_\lambda \Gamma h_v^{(b)}= \left[\sigma_1^{(c)}(\w) v_\mu v_\lambda +\sigma_2^{(c)}(\w) v_\mu \vp_\lambda + \sigma_3^{(c)}(\w) g_{\mu\lambda}\right]\bar{\Psi}_{\vp}^{\mu\,(c)}\Gamma \Psi_v^{(b)}.
\eea
with $\Psi_{\vp}^{\mu\,(c)}$ and $\Psi_v^{(b)}$ the HQET fields which destroy the spin 1/2 and spin 3/2 members of the charm  $j_{\rm ldof}^P=1^-$ doublet and the ground state $\lambb$, respectively.
Multiplying Eq.~\eqref{eq:Dterm} by $v^{\prime\lambda}$ and making use of the equation of motion  $(\vp\cdot D)h_{\vp}^{(c)}=0$, one obtains $\sigma_2^{(c)}(\w)=-\w \sigma_1^{(c)}(\w)$. Additionally, translational invariance allows to write~\cite{Leibovich:1997az}
\be
\sigma_3^{(c)}(\w) = ({\bar\Lambda}-\w{\bar\Lambda}^\prime)\sigma(\w)+(\w^2-1)\sigma_1^{(c)}(\w)
\ee
Then one has
\bea
\langle \Lambda_{c}^*;j_{\rm ldof}^P=1^-|\bar{h}_{\vp}^{(c)}i\overleftarrow{\slashed{D}} \Gamma h_v^{(b)} |\Lambda_b\rangle = \bar{\mathcal{U}}_c^\mu (\vp) \left[v_\mu \left( \slashed{v}-\w\right) \sigma_1^{(c)}(\w) +\gamma_\mu \sigma_3^{(c)}(\w) \right]{ \Gamma} u_b(v),
\eea
and thus 
\bea\label{eq:nlo:current}
\langle \lcg |\bar{h}_{\vp}^{(c)}i\overleftarrow{\slashed{D}} \Gamma h_v^{(b)}|\lambb\rangle \al = \al \frac{1}{\sqrt{3}}\bar{u}_c\left[(\w^2-1) \sigma_1^{(c)}(\w)-3\sigma_3^{(c)}(\w)\right] \gamma_5\Gamma u_b, \nno
\langle \lce |\bar{h}_{\vp}^{(c)}i\overleftarrow{\slashed{D}} \Gamma h_v^{(b)}|\lambb\rangle \al = \al  \sigma_1^{(c)}(\w) v_\lambda \bar{u}_c^\lambda (\slashed{v}-\w) \Gamma u_b.
\eea

In addition, the correction $\phi_\text{kin}^{(c)}(\w)$  from the charm kinetic energy operator $\bar{h}_{\vp}^{(c)}(iD_\perp)^2h_{\vp}^{(c)}$ in Eq.~\eqref{eq:lagO} respects HQSS and hence it enters in the same way as the leading IW function $\sigma(\w)$,  and it simply re-normalizes the latter \cite{Roberts:1992xm,Leibovich:1997az}
\be
\sigma(\w)\to \widetilde\sigma(\w) = \sigma(\w) + \frac{\phi_\text{kin}^{(c)}(\w)}{2m_c}
\ee
However, the chromomagnetic operator $\bar{h}_{\vp}^{(c)}g_s\sigma\cdot G h_{\vp}^{(c)}/2$ breaks HQSS and it leads to a contribution to the matrix elements of $\bar{h}_{v^\prime}^{(c)}\Gamma h_v^{(b)}$ \cite{Roberts:1992xm,Leibovich:1997az}
\bea\label{eq:nlo:lag}
\al \al i \phi_\text{mag}^{(c)}\,g_{\mu\alpha}\, v_\nu\,\bar{\mathcal{U}}_c^\alpha(\vp) \sigma^{\mu\nu} \frac{1+\slashed{v}^\prime}{2}\Gamma u_b(v) =  \phi_\text{mag}^{(c)} v_\lambda \bar{u}_c^\lambda(\vp) \Gamma u_b(v)-\frac{2\phi_\text{mag}^{(c)}}{\sqrt{3}}\bar{u}_c(\vp)\left(\slashed{v}-\w\right)\gamma_5 \Gamma u_b(v)
\eea
Combining the leading order (Eq.~\eqref{eq:lo2}), and the $\mathcal{O}(\Lambda_\text{QCD}/m_c)$ (Eqs.~\eqref{eq:nlo:current}-\eqref{eq:nlo:lag}) contributions, we obtain that the general form of the semileptonic matrix elements, keeping the invariance under spin rotations of the quark $b$, reads~\cite{Nieves:2019kdh},
\bea\label{eq:ff:hqss}
\langle \lcg | \bar{c}\Gamma b|\lambb\rangle \al = \al \frac{1}{\sqrt{3}}\bar{u}_c\left[ (\slashed{v}-\w)\Delta_1(\w) -\Delta_2(\w)\right]\gamma_5\Gamma u_b, \nno 
\langle \lce | \bar{c}\Gamma b| \lambb\rangle \al = \al v_\lambda \bar{u}_c^\lambda \left[ \Omega_1(\w) -(\slashed{v}-\w)\Omega_2(\w)\right]\Gamma u_b, 
\eea
where we have introduced the scalar form-factors
\bea\label{eq:deltaomega}
\Delta_1(\w) \al  = \al \sigma(\w) +\frac{1}{2m_c}\left [\phi_\text{kin}^{(c)}(\w) -2\phi_\text{mag}^{(c)}(\w)\right] = \widetilde\sigma(\w) -\frac{\phi_\text{mag}^{(c)}(\w)}{m_c},\nno \nno
\Delta_2(\w) \al = \al \frac{(\w^2-1)\sigma_1^{(c)}(\w)-3\sigma_3^{(c)}(\w)}{2m_c}=-\frac{2(\w^2-1)\sigma_1^{(c)}(\w)+3({\bar\Lambda}-\w{\bar\Lambda}^\prime)\sigma(\w)}{2m_c},\nno \nno
\Omega_1(\w) \al = \al \sigma(\w) + \frac{1}{2m_c}\left[\phi_\text{kin}^{c}(\w) + \phi_\text{mag}^{(c)}(\w)\right] =  \widetilde\sigma(\w) + \frac{\phi_\text{mag}^{(c)}(\w)}{2m_c}, \qquad 
\Omega_2(\w) = \frac{\sigma_1^{(c)}(\w)}{2m_c}.
\eea
which are determined by the leading IW function $\sigma$ and the sub-leading $\mathcal{O}(\Lambda_\text{QCD}/m_c)$ correction discussed in Ref.~\cite{Leibovich:1997az}.

 It is worth stressing that the $\mathcal{O}(\Lambda_\text{QCD}/m_c)$  decomposition of Eq.~\eqref{eq:ff:hqss} is valid   for any $\lambb$ transition  to $J^P=\frac12^-$ and $J^P=\frac32^-$ baryons, regardless of  the ldof quantum numbers in the final charm states. However the expressions of the $\Delta_{1,2}$ and $\Omega_{1,2}$ form-factors  in terms of the leading and sub-leading IW functions  are specific to $\lambb$ decays into the two members of the $j_{\rm ldof}^P=1^-$ HQSS doublet.  Otherwise, the relations of Eq.~\eqref{eq:deltaomega} are lost and thus, for instance, $\Delta_1(\w)$ does not need to approach  $\Omega_1(\w)$ in the heavy  charm quark mass limit. 
This might be of special interest, since as we argued in the introduction, the $\lamcg$ [$J^P=\frac12^-$] could contain, in addition to $j_{\rm ldof}^P=1^-$,   ldof components with $j_{\rm ldof}^P=0^-$ quantum numbers. For the case of this latter unnatural transition, the
matrix elements of the $1/m_Q$ current and kinetic energy
operator corrections are zero for the same reason that
the leading form factor vanished~\cite{Leibovich:1997az}. The time ordered
products involving the chromomagnetic operator lead to
non-zero contributions, which however vanish at zero
recoil \cite{Leibovich:1997az} and can be cast in a $\Delta_1$-type form factor.
To order $1/m_Q$, the corresponding $\Delta_2$ form-factor is therefore zero.  These different HQSS  patterns for the $\lambb$ semileptonic decay into a charm $j_{\rm ldof}^P=1^-$  doublet or a $j_{\rm ldof}^P=0^-$  singlet might shed light into
the ldof configuration in the $\lamcg$ resonance. Similarly, the $J^P=\frac32^-$ $\lamce$ resonance could have, in addition to $j_{\rm ldof}^P=1^-$, a $j_{\rm ldof}^P=2^-$ component. For this also unnatural $0^+\to 2^-$ transition, in contrast to what we saw  for the $j^P_{\rm ldof}=1^-$ component, the leading IW function and the kinetic energy operator correction vanish, as it occurs for the  $0^+\to 0^-$ case~\cite{Roberts:1992xm,Leibovich:1997az}.  
Different to the latter case, the matrix element of the $1/m_Q$ current term does not vanish  for the transition to $j_{\rm ldof}^P=2^-$, and  it takes a form that can be absorbed into the $\Omega_2(\w)$ form factor. The chromomagnetic correction provides a $1/m_c-$suppressed  contribution to $\Omega_1(\w)$. Both corrections, however, vanish at zero recoil.

By employing Eq.~\eqref{eq:ff:hqss}, it is straightforward to obtain for $\lcg$ 
\bea\label{eq:ff:HQET12}
\langle \lcg | \bar{c} b|\lambb\rangle \al = \al -\frac{1}{\sqrt{3}} \bar{u}_c \left[ (\w+1)\Delta_1 + \Delta_2 \right] \gamma_5u_b, \nno
\langle \lcg | \bar{c}\gamma_5 b|\lambb\rangle \al = \al -\frac{1}{\sqrt{3}} \bar{u}_c\left[ (\w-1)\Delta_1 + \Delta_2 \right] u_b, \nno 
\langle \lcg | \bar{c}\gamma_\mu b|\lambb\rangle \al = \al \frac{1}{\sqrt{3}} \bar{u}_c \big( [(\w-1)\Delta_1 +\Delta_2]\gamma_\mu -2 \Delta_1 v_\mu \big)\gamma_5 u_b, \nno 
\langle \lcg | \bar{c}\gamma_\mu \gamma_5 b|\lambb\rangle \al = \al \frac{1}{\sqrt{3}} \bar{u}_c \big( [(\w+1) \Delta_1 +\Delta_2]\gamma_\mu - 2\Delta_1 v_\mu\big)u_b,\nno 
\langle \lcg | \bar{c}\sigma_{\mu\nu}\gamma_5 b|\lambb\rangle \al = \al -\frac{1}{\sqrt{3}} \bar{u}_c \big( [\Delta_1(\w-1)+\Delta_2 ] \sigma_{\mu\nu} + 2i \Delta_1(\gamma_\mu v_\nu - \gamma_\nu v_\mu)\big) u_b,
\eea
and for $\lce$,
\bea\label{eq:ff:HQET32}
\langle \lce | \bar{c} b|\lambb\rangle \al = \al \bar{u}_c^\lambda v_\lambda [\Omega_1+(\w-1)\Omega_2 ] u_b, \nno 
\langle \lce | \bar{c} \gamma_5 b|\lambb\rangle \al = \al \bar{u}_c^\lambda v_\lambda [\Omega_1+(\w+1)\Omega_2 ] u_b, \nno
\langle \lce | \bar{c} \gamma_\mu b|\lambb\rangle \al = \al \bar{u}_c^\lambda v_\lambda \big( [\Omega_1 + (\w+1) \Omega_2 ]\gamma_\mu -2\Omega_2 v_\mu \big) u_b, \nno
\langle \lce | \bar{c} \gamma_\mu\gamma_5 b|\lambb\rangle \al = \al \bar{u}_c^\lambda v_\lambda \big( [\Omega_1 + (\w-1)\Omega_2] \gamma_\mu -2 \Omega_2 v_\mu \big) \gamma_5 u_b, \nno 
\langle \lce | \bar{c} \sigma_{\mu\nu} b|\lambb\rangle \al = \al \bar{u}_c^\lambda v_\lambda \big([\Omega_1 + (\w-1)\Omega_2]\sigma_{\mu\nu} + 2i \Omega_2 (\gamma_\mu v_\nu - \gamma_\nu v_\mu ) \big) u_b.
\eea
The matrix elements above can be used to express  the form factors $d_i$ and $l_i$ in terms of  $\Delta_{1,2}$ and $\Omega_{1,2}$. The relations are collected in  Appendix~\ref{app:ff:hqss}.

The  $\lambb\to\lcg \ell \bar\nu_\ell$ and $\lambb\to\lce \ell \bar\nu_\ell$ SM differential decay widths deduced from the vector and axial matrix elements in Eqs.~(\ref{eq:ff:HQET12}) and ~(\ref{eq:ff:HQET32}) are  given by~\cite{Nieves:2019kdh}
\begin{eqnarray}
\frac{d\Gamma[\Lambda_b\to \Lambda_{c,J^P}^*]}{d\omega}&=& \left(2J+1\right)\frac{8\,\Gamma_0}{3}\left(\frac{m_{\Lambda_c^*}}{m_{\Lambda_b}}\right)^3
\left(1-\frac{m^2_\ell}{q^2}\right)^2(\omega^2-1)^J\,
\left\{ \alpha_J^2\Bigg[3\omega \frac{q^2+m_\ell^2}{m_{\Lambda_b}^2} \right. \nonumber \\
&& \left.+ 2\frac{m_{\Lambda_c^*}}{m_{\Lambda_b}}(\omega^2-1)
\left(1+\frac{2m^2_\ell}{q^2}\right)\Bigg]+2\,(\omega^2-1)\,\left[\alpha_1(\omega)\alpha_2(\omega)\right]_J \times\right.\nonumber \\
&&\times \left. \left[\frac{2q^2+ m_\ell^2}{m_{\Lambda_b}^2} +\left(1
-\frac{m^2_{\Lambda_c^*}}{m_{\Lambda_b}^2}\right)\left(1+\frac{2m^2_\ell}{q^2}\right)\right]+ \mathcal{O}\left(\frac {\Lambda_\text{QCD}}{m_b}\right)\right\}\label{eq:gamma_diff}
\end{eqnarray}
with $m_\ell$ the final charged lepton mass, $\Gamma_0 =  |V_{cb}|^2 G_F^{\,2}m_{\Lambda_b}^5/(192\pi^3)$, where $|V_{cb}|$ is the modulus of the Cabibbo--Kobayashi--Maskawa  matrix element for the $b\to c$ transition and $G_F= 1.16638 \times \
10^{-11}$\,MeV$^{-2}$ is the Fermi decay constant, and
\begin{eqnarray}
 \alpha_{J=1/2}^2(\omega) &=& \Delta_2^2(\omega)+(\omega^2-1)\Delta_1^2(\omega), \qquad \left. \alpha_{1}(\omega)\,\alpha_{2}(\omega)\,
 \right|_{J=1/2} =\Delta_1(\omega)\,\,\Delta_2 (\omega)\,\nonumber\\
 \alpha_{J=3/2}^2(\omega) &=& \Omega_1^2(\omega)+(\omega^2-1)\Omega_2^2(\omega), \qquad \left. \alpha_{1}(\omega)\,\alpha_{2}(\omega)\,
 \right|_{J=3/2} =\Omega_1(\omega)\,\,\Omega_2(\omega)\label{eq:gamma_diff2}
\end{eqnarray}
In the strict $m_c\to \infty$ limit,
\begin{eqnarray}
\lim _{m_c\to \infty}\left(\frac{d\Gamma[\Lambda_b\to \lcg]/d\omega}{d\Gamma[\Lambda_b\to \lce]/d\omega}\right)&=& \frac12 \, \left(\frac{\lim _{m_c\to \infty}\Delta_1(\omega)}{\lim _{m_c\to \infty}\Omega_1(\omega)}\right)^2 \label{eq:HQratio}
\end{eqnarray}
and if $\lcg$ and $\lce$ are the two members of the $j_{\rm ldof}^P=1^-$ HQSS doublet, the above ratio will be 1/2, since  the relations of Eq.~\eqref{eq:deltaomega}  will be satisfied and  $\Delta_1(\omega)\sim \Omega_1(\omega) \sim \sigma(\omega)$. However in the vicinity of zero recoil ($\w \le 1.05$), one might find large deviations, because in this kinematic region, the sub-leading contribution  $\Delta_2^2(\omega)$  to $\alpha_{J=1/2}^2(\omega)$ could be comparable or larger than the other one, $(\omega^2-1)\Delta_1^2(\omega)$, which is used to obtain\footnote{Note that $(\omega^2-1)$ multiplying $\Delta_1^2(\omega)$ compensates the overall extra  $(\omega^2-1)$ factor which appears for the $J^P=3/2^-$ decay width in Eq.~\eqref{eq:gamma_diff}.  } Eq.~\eqref{eq:HQratio}. Actually, the $\Delta_2(\omega)$ form-factor accounts for a S-wave term to the $\lambb\to \lcg$ decay amplitude proportional to  $3({\bar\Lambda}-{\bar\Lambda}^\prime)\sigma(1)/(2m_c) \sim(0.2-0.3) \sigma(1)$~\cite{Nieves:2019kdh}. This follows from  Eq.~\eqref{eq:deltaomega}, using $m_c\sim 1.4$ GeV for the charm quark mass, $\bar\Lambda\sim 0.8$ GeV and $\bar\Lambda'\sim 1 \pm 0.1 $ GeV for 
the energies of the ldof, in the $m_Q\to \infty$ limit, in the $\Lambda_b$ and the P-wave $\Lambda_c^*\,(j_q^P=1^-)$ baryon. Since the $\lambb\to \lce$ semileptonic decay proceeds necessarily in P-wave, it appears an extra factor $(\omega^2-1)$, and the ratio $\frac{d\Gamma[\Lambda_b\to \lcg]/d\omega}{d\Gamma[\Lambda_b\to \lce]/d\omega}$ could differ significantly from $1/2$ when $\omega$ is close to 1. 

\section{Numerical analysis}\label{sec:num}

In the last two sections, we have presented a general parameterization  and $\mathcal{O}(\Lambda_\text{QCD}/m_c)$ HQET form factors for the $\lambb\to \lcg/\lce$ semileptonic transitions. Although in recent years, the LHC has collected large numbers of $\lambb\to\lamcg \mu^-\bar{\nu}_\mu$ and $\lambb\to \lamce\mu^-\bar{\nu}_\mu$ samples~\cite{LHCb:2017vhq}, the extraction of the form factors is, however, not available yet. Without  experimental input for the form factors, LQCD provides a valuable framework to test how well the  $j_{\rm ldof}^P=1^-$  HQSS predictions, with corrections of order  $\omc$, works to describe the $\lamcg$ and $\lamce$ form factors measured on the lattice, and whether there is a possibility of a sizable  $j_{\rm ldof}^P=0^-$ configuration for the $\lamcg$ resonance.
\subsection{LQCD  and $\omc$ HQET form-factors}
\begin{figure}
\includegraphics[width=0.8\textwidth]{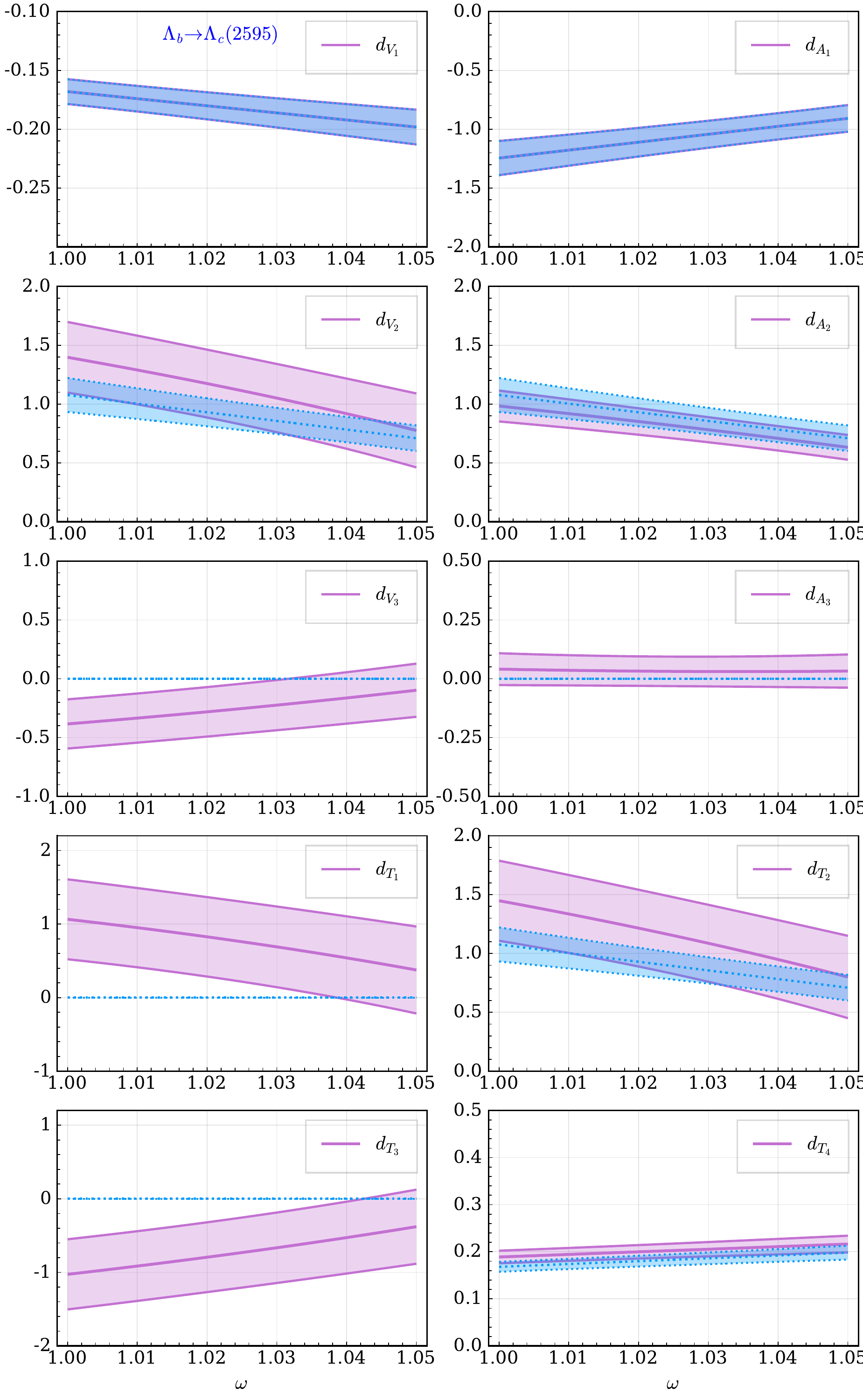}
\caption{Comparison between the LQCD~\cite{Meinel:2021mdj} (purple uncertainty bands and solid curves) and the $\omc$ HQET  (Eqs.~\eqref{eq:nDeltaOmega} and \eqref{eq:ff:HQET}) form factors for the $\lambb\to\lamcg$ transition, as a function of $\w$. The HQET predictions are displayed by blue bands and dotted lines. The LQCD results are obtained using the relations given in Appendix \ref{app:LQCDFF} between the form-factors displayed here and the helicity ones computed in Ref.~\cite{Meinel:2021mdj}.}\label{fig:ff12}
\end{figure}
\begin{figure}
\includegraphics[width=0.8\textwidth]{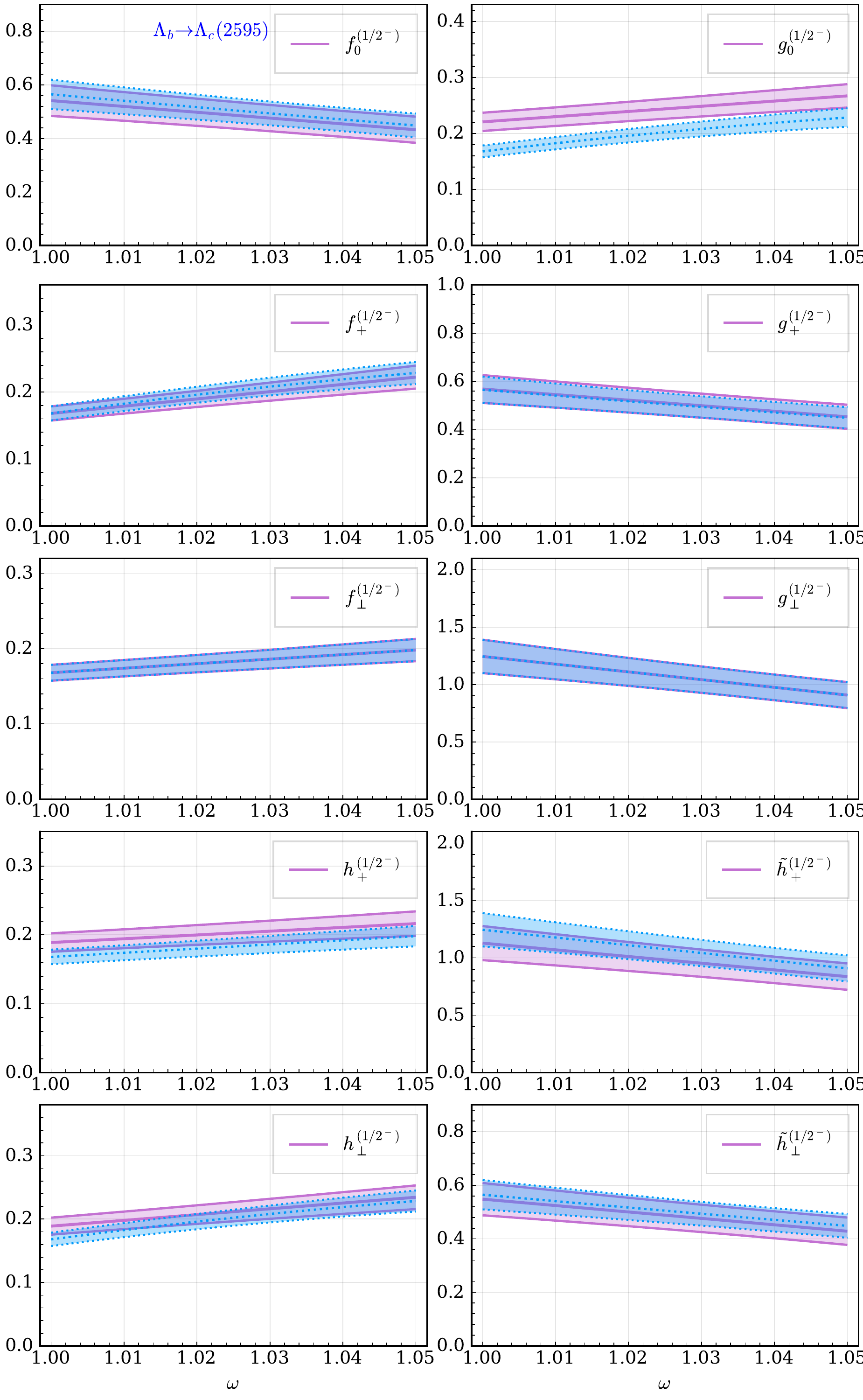}
\caption{Comparison of the $\lambb\to\lamcg$  LQCD (purple uncertainty bands and solid curves)  and $\omc$ HQET results for the  helicity form factors, which are those directly computed in Refs.~\cite{Meinel:2021rbm,Meinel:2021mdj}, as function of $\w$. The HQET predictions are obtained by inverting the relations given in Appendix~\ref{app:LQCDFF}, and using  for the $d_i$ form-factors, the  $\omc$ expressions of Eq.~\eqref{eq:ff:HQET}.  The HQET predictions are displayed by blue bands and dotted lines. }\label{fig:ff12:helicity}
\end{figure}
The first  LQCD calculation of the form factors for these transitions was carried out in Ref.~\cite{Meinel:2021rbm}, using three different ensembles of gauge-field configurations with $2+1$ flavors of domain-wall fermions generated by the RBC and UKQCD collaborations~\cite{RBC:2010qam,RBC:2014ntl}, and  three-quark interpolating fields to excite $\Lambda_c^*$ states.  The form-factors in Ref.~\cite{Meinel:2021rbm} are defined through a helicity decomposition of the amplitudes and are evaluated at three unphysical pion masses, $m_\pi=0.4312(13)$, $0.3400(11)$ and $0.3030(12)$ GeV. Finally, the results were  extrapolated to the physical point (continuum limit and physical pion mass), where the LQCD form factors are parameterized as \cite{Meinel:2021rbm,Meinel:2021mdj},
\bea\label{eq:parameterization}
f(\w) = F^f + A^f(\w-1),
\eea
which corresponds to a Taylor expansion of the form factors around zero recoil $\w=1$. Because only two different close kinematics ($\w=1.01$ and $\w=1.03$) are available at the physical point in the LQCD calculation of Ref.~\cite{Meinel:2021rbm}, the parameterization of Eq.~\eqref{eq:parameterization} is expected to be only reliable for small ($\w-1$). Systematic uncertainties are estimated from the variation of results obtained from two different extrapolations of the lattice results to the physical limit. The helicity form factors, which parameterize  the matrix elements of the vector, axial-vector and tensor currents,  for the two available values of $\w$ are treated in Ref.~\cite{Meinel:2021rbm} as independent quantities. Thus, a total of 48  parameters were fitted to the lattice data.  
In the improved analysis carried out by the same authors in Ref.~\cite{Meinel:2021mdj}, relations among the different form factors at zero recoil, which follow from rotational symmetry, are imposed to ensure the correct behavior of the angular observables near the endpoint, and thus the number of free parameters is reduced to 39. In this work, we will employ the updated results of Ref.~\cite{Meinel:2021mdj}. 
The form factors $d_i$ and $l_i$ introduced in Eqs.~\eqref{eq:formfactor12} and~\eqref{eq:formfactor32} are computed from the helicity form factors determined in Refs.~\cite{Meinel:2021rbm,Meinel:2021mdj} with the help of Eqs.~\eqref{eq:htod} and \eqref{eq:htol}, respectively.

As discussed in Sec.~\ref{sec:hqss}, there are only two independent functions $\Delta_{1,2}(\w)$ and $\Omega_{1,2}(\w)$ for the $\lambb\to\lcg$ and $\lambb\to\lce$ decays, respectively, including up to $\omc$ contributions. This is to say, neglecting $\mathcal{O}(\Lambda_\text{QCD}/m_b)$ and perturbative QCD short distance corrections.  In order to test if the HQSS describes the semileptonic $\lambb\to \lamcg$ and $\lambb\to\lamce$ decays at this order, we determine the $\Delta_{1,2}(\w)$ and $\Omega_{1,2}(\w)$ functions  from two of the form factors in each of the transitions, and then predict the rest of the form factors using the relations collected in Appendix~\ref{app:ff:hqss}. We take, 
\bea\label{eq:nDeltaOmega}
\al \Delta_1(\w) = \dfrac{\sqrt{3}}{2}\left[d_{A_1}(\w)-d_{V_1}(\w) \right],\quad \al \Delta_2(\w)  = \frac{\sqrt{3}}{2}\left[ (1+\w) d_{V_1}(\w)  + (1-\w) d_{A_1}(\w) \right], \nno
\al \Omega_1(\w)  = \dfrac{1}{2}\left[ (1-\w)l_{V_1}(\w) +(1+\w) l_{A_1}(\w) \right], \quad \al \Omega_2(\w) = \dfrac12 \left[ l_{V_1}(\w) - l_{A_1}(\w) \right].
\eea
Here there is a subtle point. We note that all type of contributions suppressed by any power of  the charm quark mass should be included in the {\it unknown}  $\Delta_{1,2}$ and $\Omega_{1,2}$ functions empirically determined from the LQCD form-factors. Thus, we  expect only perturbative QCD short distance~\cite{Papucci:2021pmj} and $\mathcal{O}(\Lambda_\text{QCD}/m_b)$ corrections, from the breaking of $b$-quark spin rotational invariance, to the  matrix elements of the scalar, pseudoescalar, vector, axial-vector and tensor currents between the $\lambb$ ground state and the final odd parity charm  $\lamcg$ and $\lamce$ resonances,  for the whole range of $\omega$ values accessible in the decay.  However, far from zero recoil, the energy scale of the heavy quark symmetry breaking corrections will not  necessarily be  $\Lambda_{\rm QCD}\sim (200-300)$ MeV anymore. 
For instance,  the momentum of the outgoing hadron in the $\lambb$-rest frame  ($|\vec{q}\,|= m_{\Lambda_c^*}\sqrt{\omega^2-1}\,$) could be much larger than $\Lambda_{\rm QCD}$. For $\omega=1.05$, we have already $|\vec{q}\,|/m_c > 1/2$ and the $(\Delta_1,\Delta_2)$ or $(\Omega_1,\Omega_2)$ hierarchy relations inferred from Eqs.~\eqref{eq:deltaomega} clearly break down.  Even corrections suppressed by the bottom quark mass might become relevant for $\omega\ge 1.1$, where  $|\vec{q}\,| \ge 1.2$ GeV. Note that $\omega$ reaches values of about $\sim 1.2\, (1.3)$ at the end-point of the spectrum for the tau (muon) semileptonic mode.    

The  $\lambb\to\lamcg$  and $\lambb\to\lamce$ form factors $d_i$ and $l_i$ evaluated using Eq.~\eqref{eq:ff:HQET} are compared with the LQCD results in Figs.~\ref{fig:ff12} ($1/2^+ \to 1/2^-)$,  \ref{fig:ff32va} (vector and axial-vector $1/2^+ \to 3/2^-$) and \ref{fig:ff32T} (tensor $1/2^+ \to 3/2^-$). There, the purple solid and blue dotted curves correspond to the LQCD results and the HQET predictions, obtained using the $\Delta_{1,2}$ or $\Omega_{1,2}$ functions determined as specified in Eq.~\eqref{eq:nDeltaOmega}, respectively. Statistic and systematic uncertainties, given in Refs.~\cite{Meinel:2021rbm,Meinel:2021mdj}, are added in quadrature and shown in the plots. 
We do not show the scalar and pseudoscalar form factors ($d_{S,P}$ and $l_{S,P}$), since they were not computed in the LQCD simulation of Refs.~\cite{Meinel:2021rbm,Meinel:2021mdj}. They are given in Appendix \ref{app:LQCDFF} using their relation to the vector and axial-vector form factors obtained from the heavy quarks equations of motion.
\begin{figure}
\includegraphics[width=0.8\textwidth]{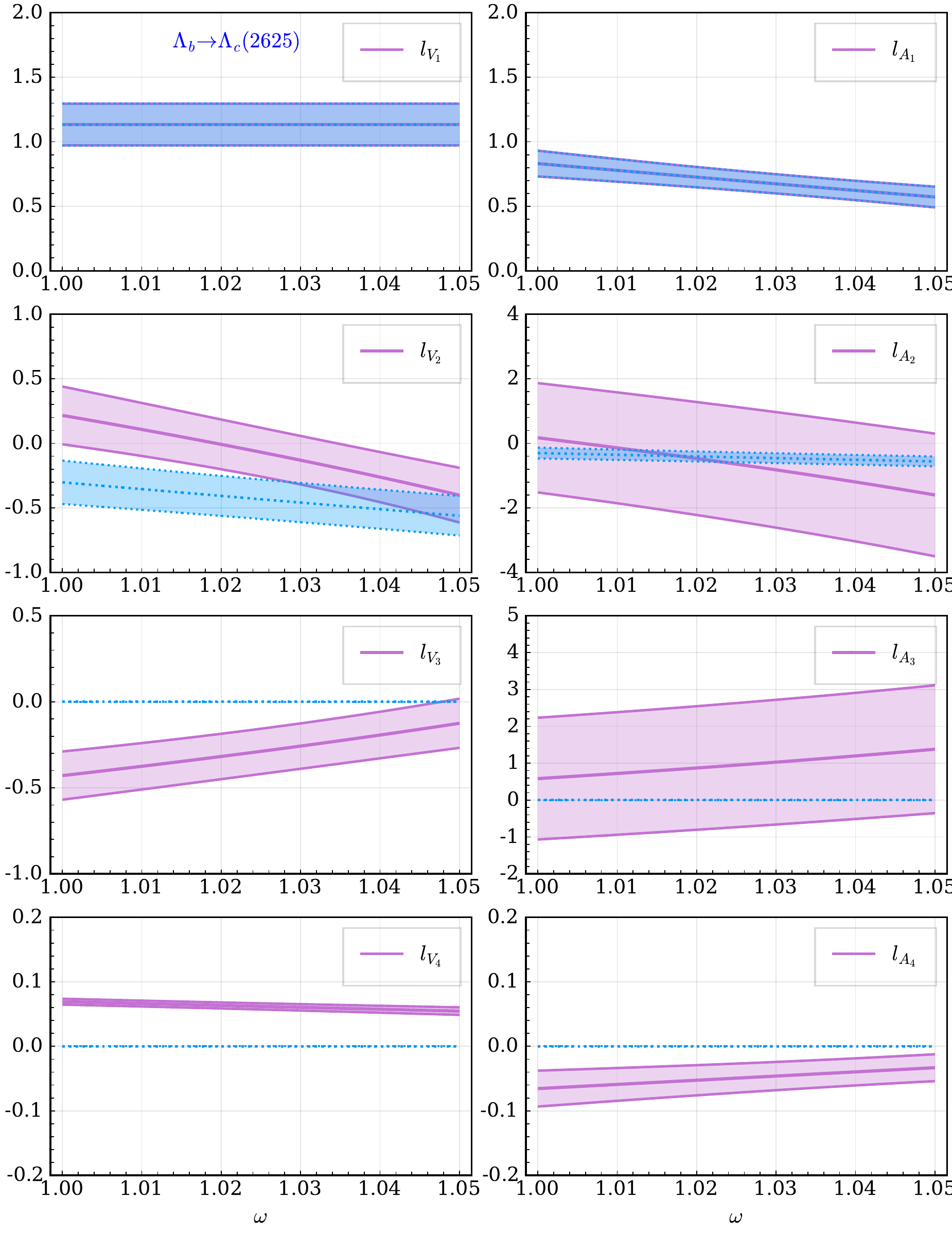}
\caption{Same as Fig.~\ref{fig:ff12} for the $\lambb\to\lamce$  vector and axial-vector form-factors. }\label{fig:ff32va}
\end{figure}
\begin{figure}
\includegraphics[width=0.8\textwidth]{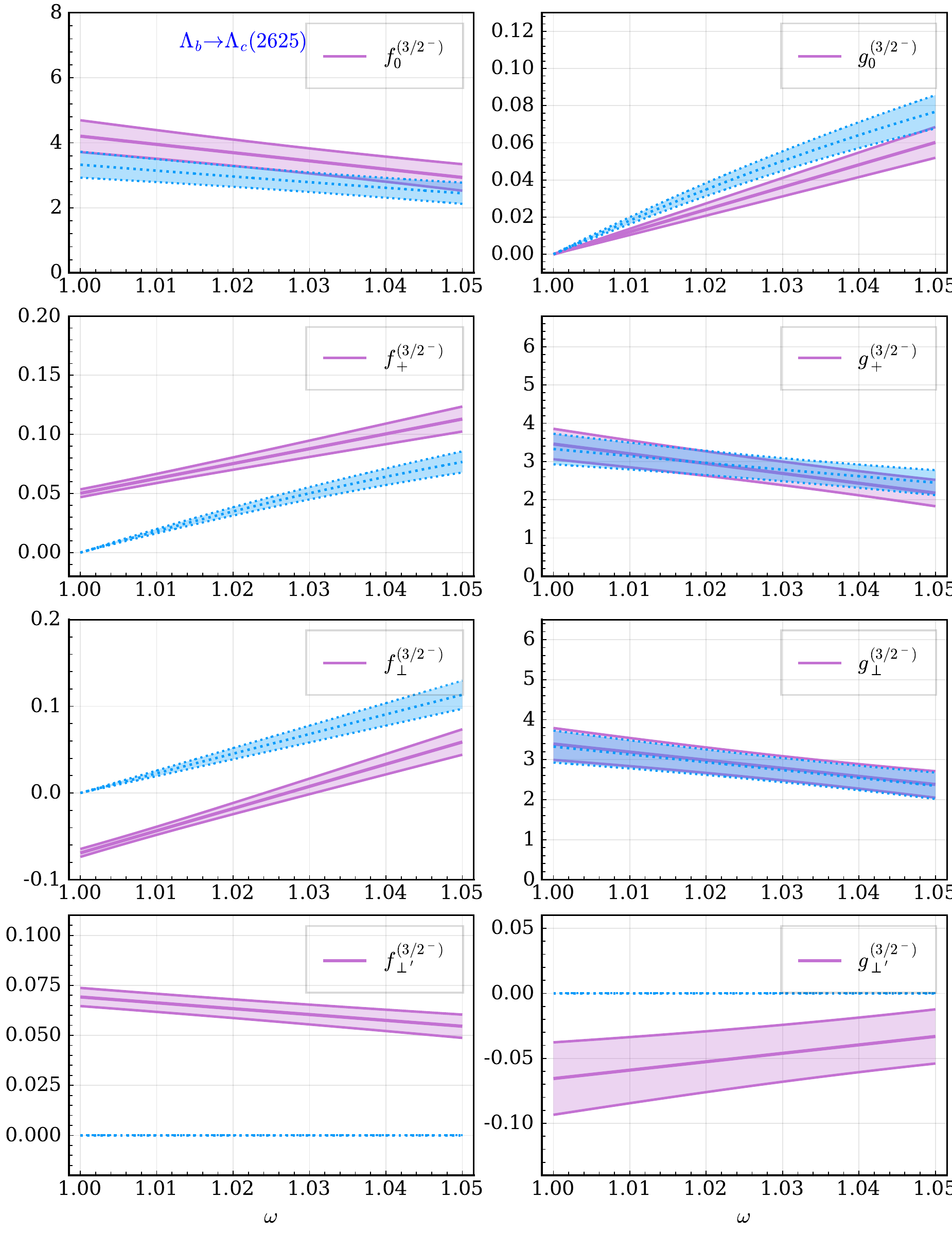}
\caption{Same as Fig.~\ref{fig:ff12:helicity} for the $\lambb\to\lamce$ scalar, pseudoscalar,vector and axial-vector helicity form-factors.}\label{fig:ff32va:helicity}
\end{figure}

\begin{figure}
\includegraphics[width=0.8\textwidth]{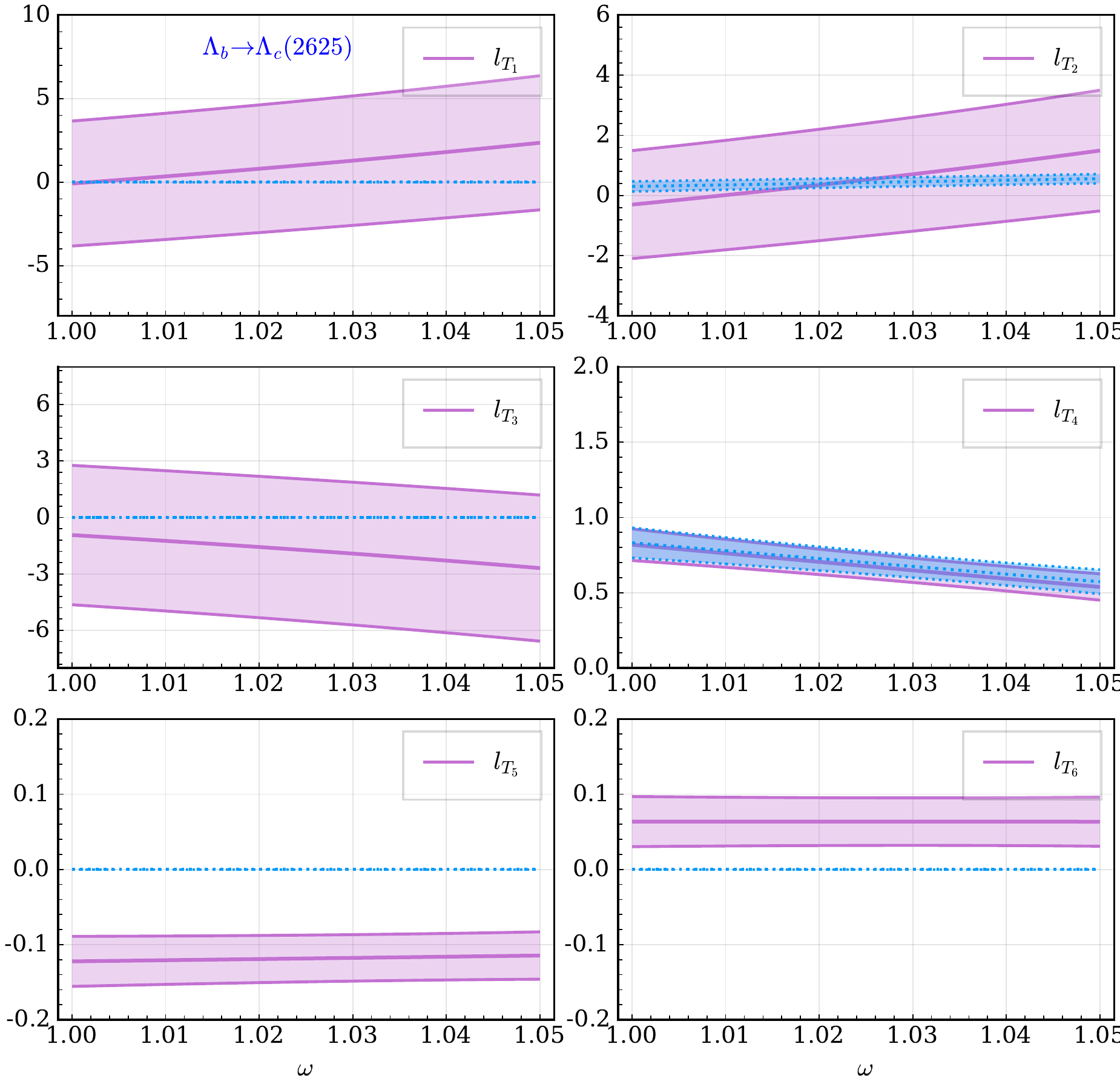}
\caption{Same as Fig.~\ref{fig:ff12} for the $\lambb\to\lamce$ tensor form-factors. }\label{fig:ff32T}
\end{figure}
\begin{figure}
\includegraphics[width=0.8\textwidth]{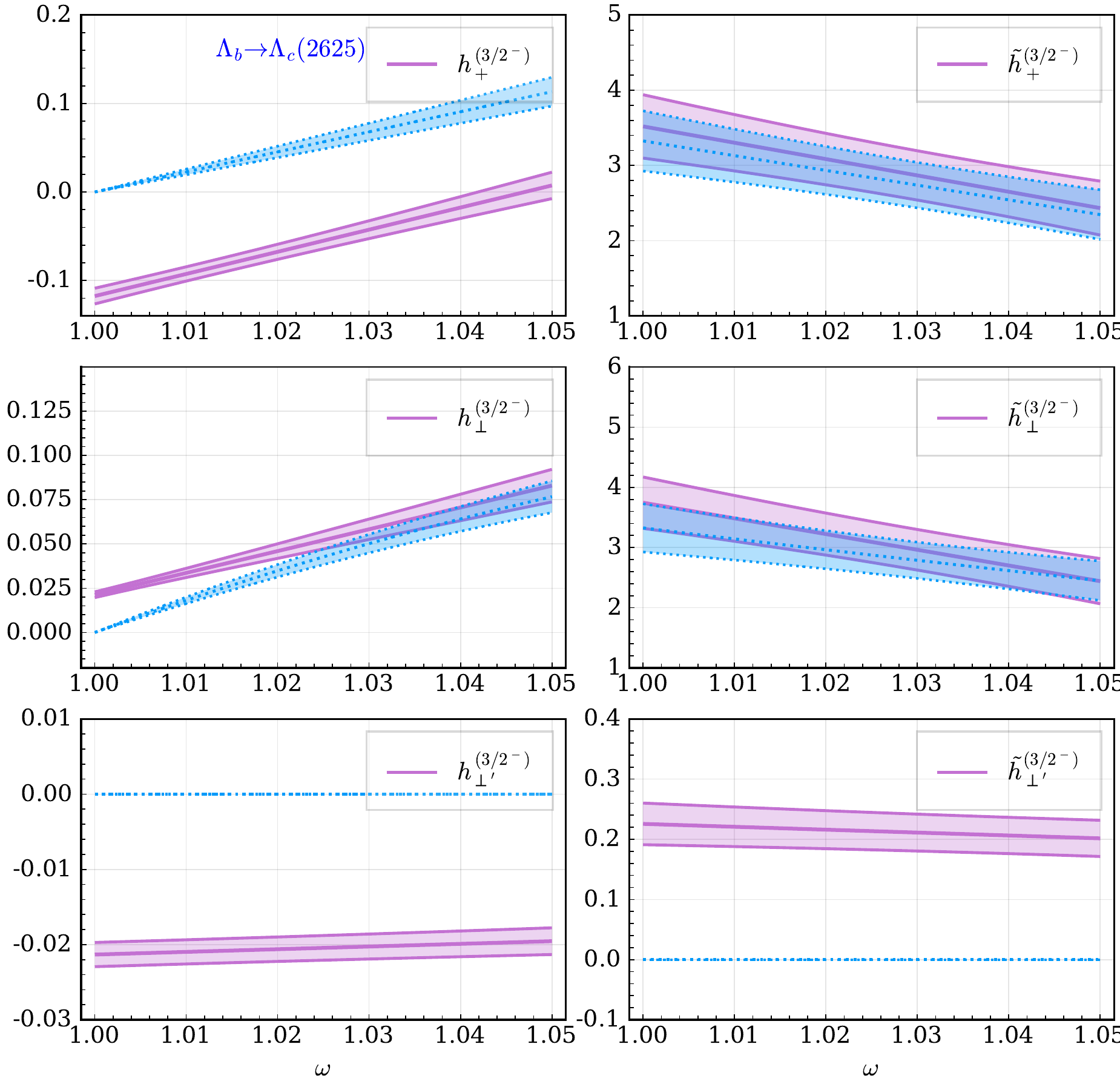}
\caption{Same as Fig.~\ref{fig:ff12:helicity} for the $\lambb\to\lamce$ tensor helicity form-factors.} \label{fig:ff32T:helicity}
\end{figure}

The comparison of the LQCD and $\omc$ HQET results for the helicity form factors, which are those directly obtained in the lattice calculation~\cite{Meinel:2021rbm,Meinel:2021mdj},  are presented in Fig.~\ref{fig:ff12:helicity} for $\lambb\to\lamcg$, and Figs.~\ref{fig:ff32va:helicity} and \ref{fig:ff32T:helicity} for $\lambb\to\lamce$. The comparison for the scalar and pseudoscalar form factors ($d_{S,P}$ and $l_{S,P}$) can be inferred from the results shown  in these figures for $f_0^{(J^P)}$ and $g_0^{(J^P)}$, cf. Eqs.~\eqref{eq:htod} and \eqref{eq:htol}.

For the $\lambb\to\lamcg$ transition, we see that LQCD and HQET agree in general  within uncertainties. The largest discrepancies are found for the vector $d_{V_3}$ and tensor $d_{T_1}$ and $d_{T_3}$ form-factors, which vanish at this order within the HQET scheme, and moreover they are poorly determined on the lattice. In fact, the large uncertainties affecting these form factors make these discrepancies of little significance. Note that the axial $d_{A_3}$ form-factor, which  also vanishes at this order of the HQET expansion,  is accurately determined  in Ref.~\cite{Meinel:2021mdj}, and it turns out to be compatible with zero in the whole  $\omega$ interval shown in the figure. The results in Fig.~\ref{fig:ff12:helicity}, where we pay attention to the helicity form-factors directly determined in the LQCD simulation, confirm the quite reasonable comparison between LQCD and HQET predictions. The most significant discrepancy is found now for the axial $\gzero$ helicity form-factor, which  contributes to the axial $d_{A_2}$ and $d_{A_3}$, where however LQCD and HQET predictions agree within errors. On the other hand, we do not appreciate significant discrepancies between both sets of predictions for the helicity form-factors involved in the vector $d_{V_3}$ and tensor $d_{T_1}$ and $d_{T_3}$.  Therefore, the origin of the differences noted above must be sought in the large cancellations responsible for these form-factors being zero at order $\omc$.

At first sight, the comparison of LQCD and $\omc$ HQET form-factors for the $\lambb\to\lamce$ decay  does not look as satisfactory as that described above for $\lambb\to\lamcg$, as one might infer from Figs.~\ref{fig:ff32va} and \ref{fig:ff32T}.
In particular, the discrepancies are clearly visible in $l_{V_3}$, $l_{V_4}$, $l_{A_4}$, $l_{T_5}$ and $l_{T_6}$, which are predicted to be zero at this order of the HQET expansion. We however note, the LQCD predictions of Ref.~\cite{Meinel:2021mdj}  for these form factors are compatible with typical values of order $\mathcal{O}(\Lambda_\text{QCD}/m_b)\sim 0.1$, except for the case of $|l_{V_3}|$, which takes notably higher values. As in the case of $d_{V_3}$ for the $\lambb\to\lamcg$ mode, we also believe that the origin of the differences for  $l_{V_3}$ stems from some inaccuracies in  the required large cancellations, among the $f_0^{(\frac32^-)}, f_+^{(\frac32^-)}, f_\perp^{(\frac32^-)}$ and $f_{\perp^\prime}^{(\frac32^-)}$ helicity form-factors,  to make this form-factor  vanish at order $\omc$. The same helicity form-factors, but in a different linear combination, appear also  in $l_{V_2}$, for which some disagreement, near zero recoil, between LQCD and  $\omc$ HQET predictions can be appreciated in Fig.~\ref{fig:ff32va}. Paying now attention to the results in Figs.~\ref{fig:ff32va:helicity} and \ref{fig:ff32T:helicity}, we observe there are five dominant helicity transitions, those associated to $f_0^{(\frac32^-)},g_+^{(\frac32^-)}$ and $g_{\perp}^{(\frac32^-)}$ for vector and axial-vector currents, and 
$ \hh_+^{(\frac32^-)}$ and  $\hh_\perp^{(\frac32^-)}$ for the tensor ones. Importantly,  the $\mathcal{O}(\Lambda_\text{QCD}/m_c)$ HQET scheme provides a good reproduction of the LQCD results for these leading form-factors. The most significant differences appear for $f_0^{(\frac32^-)}$, where the central values of both schemes are separated by one sigma, which might explain the discrepancies pointed out above for $l_{V_2}$ and $l_{V_3}$. The rest of the helicity form-factors are very small, around a factor 20-40 smaller (in absolute value) than the five dominant ones mentioned above.  Order $\mathcal{O}(\Lambda_\text{QCD}/m_b)$ and/or of short physics corrections to the HQET results or further systematic errors affecting the LQCD ones  might significantly change the predictions displayed in  Figs.~\ref{fig:ff32va:helicity} and \ref{fig:ff32T:helicity}, and improve the apparent disagreement exhibited there for these sub-leading form-factors.

Therefore, we conclude that $\omc$ HQET scheme describes reasonably well the LQCD results, taking into account that the neglected HQET sub-leading corrections or LQCD systematic uncertainties might be important in those cases where discrepancies are more apparent on a naive visual inspection of Figs.~\ref{fig:ff32va} and \ref{fig:ff32T}. The results of the next subsection will give further support to this general conclusion.

\subsection{HQSS form-factors and differential decay rates}
\begin{figure}
\includegraphics[width=0.8\textwidth]{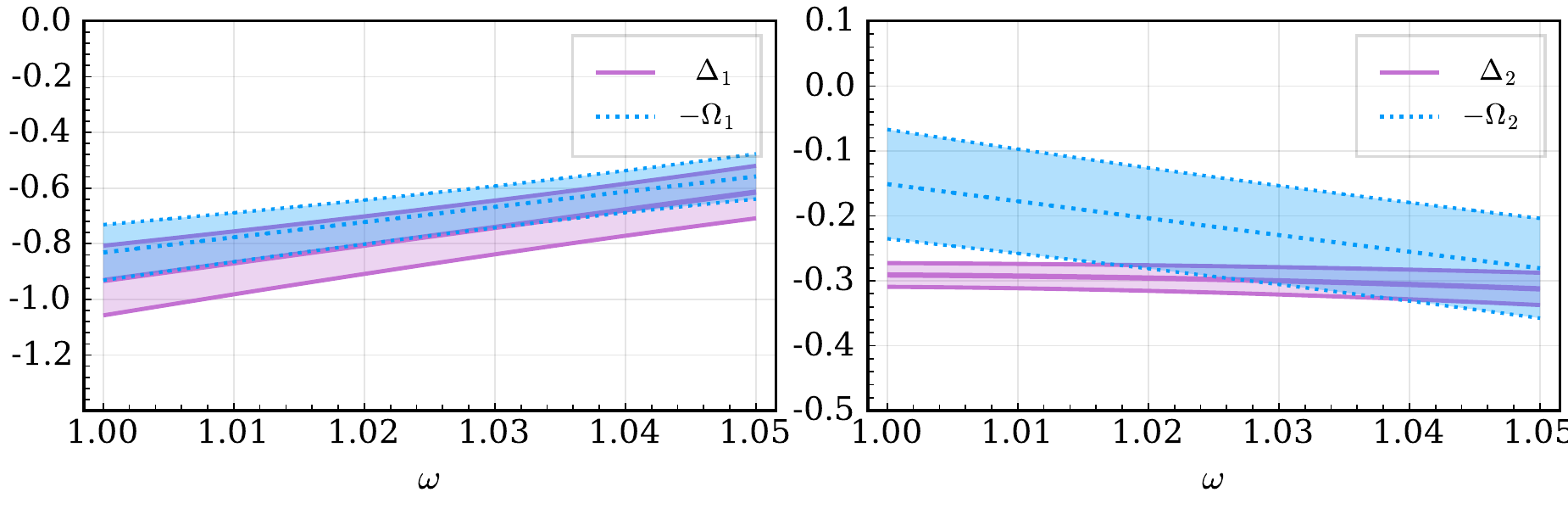}\\
\includegraphics[width=0.8\textwidth]{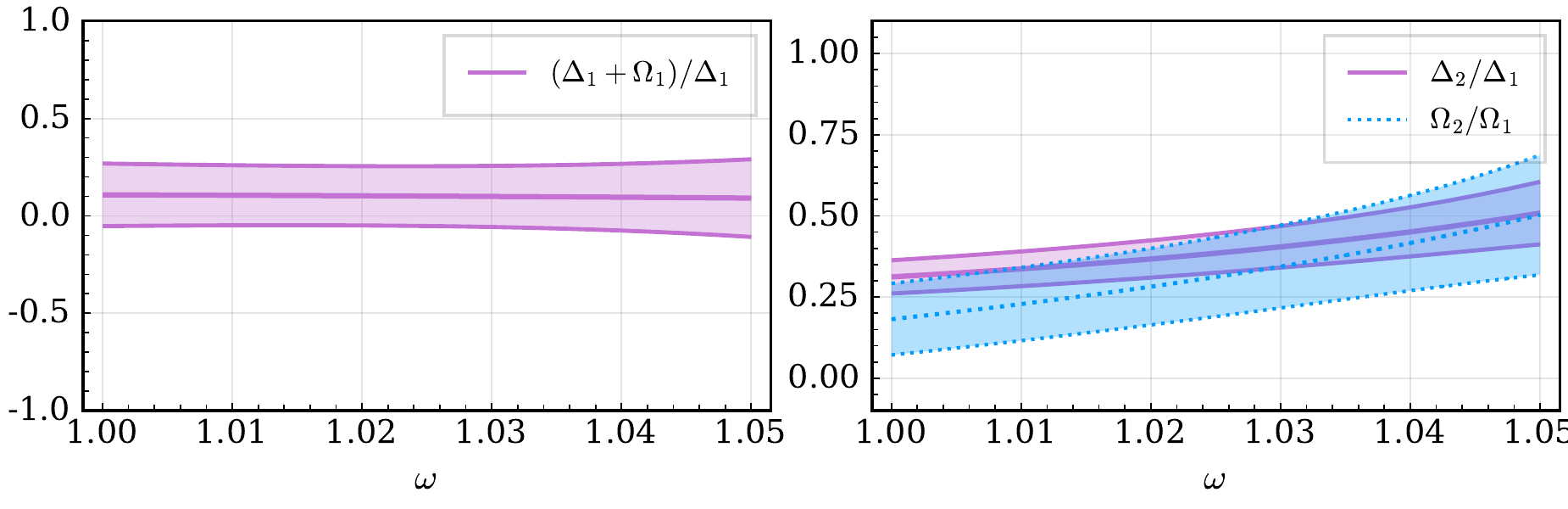}
\caption{Comparison between the leading and sub-leading $\Delta_{1,2}$ and $\Omega_{1,2}$ functions, determined as specified in Eq.~\eqref{eq:nDeltaOmega}. Note that it would be necessary to include a relative phase between the states $\lamcg$ and $\lamce$ to obtain $\Delta_1$ and $\Omega_1$ with the same sign.}\label{fig:delta_vs_omega}
\end{figure}

 In this subsection, we try first to assess how well the $\lamcg$ and $\lamce$ are described by a  $j_{\rm ldof}^P=1^-$ configuration for the ldof. We discussed in Sec.~\ref{sec:hqss} that $\Delta_2(\w)=\Omega_2(\w)=0$ in the heavy charm quark limit, and that in addition, $\Delta_1(\w)=\Omega_1(\w)=\sigma(\w)$ if the $\lamcg$ and $\lamce$ are the members of the lowest-lying HQSS $j_{\rm ldof}^P=1^-$ doublet. Thus, the difference between the $\Delta_1(\w)$ and $\Omega_1(\w)$ should be of order $\omc$, while $\Delta_2(\w)$ and $\Omega_2(\w)$ are both  of order $\omc$, and hence suppressed compared to the leading $\Delta_1(\w)$ and $\Omega_1(\w)$ functions, cf.~\eqref{eq:deltaomega}.  Note that the overall sign of the form factors for each decay mode depends on the phase conventions of the $\Lambda_c^*$ states \cite{Meinel:2021rbm}, and  only the relative signs among the form factors for a specific mode are well determined. The comparisons between the $\Delta_{1,2}$ and $\Omega_{1,2}$ IW functions, obtained from the LQCD results for the $d_{V_1},d_{A_1}, l_{V_1}$ and $l_{A_2}$ form-factors (Eq.~\eqref{eq:nDeltaOmega}) are shown in Fig.~\ref{fig:delta_vs_omega}. We see that the differences between $\Delta_1$ and $-\Omega_1$ are small compared to any,   $\Delta_1$ or $\Omega_1$, of these leading IW functions, and can be naturally attributed to $\omc$ contributions. This gives some support, or at least does not contradict, that the  $\lamcg$ and $\lamce$ resonances  might form the lowest-lying  HQSS $j_{\rm ldof}^P=1^-$ doublet. Nevertheless, we would remind 
 that in the LQCD simulation carried out in Refs.~\cite{Meinel:2021rbm, Meinel:2021mdj}, the $\Lambda_c^*$ states were described using three-quark interpolating fields. This should not create any bias for an  unquenched simulation, since these operators should capture any baryon-meson ($\Sigma_c^{(*)}\pi$) component in the QCD state, assuming a sufficiently large evolution time. The chiral-continuum extrapolations of the masses of the states excited on the lattice are~\cite{Meinel:2021rbm}
\be
\mcg = \left (2693 \pm  43_{\rm stat} \pm 95_{\rm sys}\right)\,\, {\rm MeV}, \qquad \mce = \left (2742 \pm  43_{\rm stat} \pm 96_{\rm sys}\right)\,\, {\rm MeV} 
\ee
 which, though are consistent with the experimental values of $\mcg = 2592.25 (28)$ MeV and $\mce= 2628.11 (19)$ MeV ~\cite{ParticleDataGroup:2020ssz}, are not accurate enough to disentangle the possible effects of the $\Sigma_c \pi$ and $\Sigma_c^*\pi$ thresholds located only a few MeV above the position of the physical resonances, especially in the case of the $\lamcg$ and  $\Sigma_c \pi$ ~\cite{Nieves:2019nol}. Therefore, the claim made in this latter reference 
 that the $\lamcg$ and $\lamce$ resonances are not HQSS partners cannot be discarded using the LQCD data studied here. In Ref.~\cite{Nieves:2019nol}, the $J^P=3/2^-$ state is described  mostly as a dressed three-quark state, whose origin is determined by a bare state~\cite{Yoshida:2015tia}, predicted to lie very close to the mass of the resonance. The  $J^P=1/2^-$ resonance seemed to have, however, a predominant molecular structure. This is, depending on the renormalization scheme, either because the $\lamcg$ is the result of the chiral $\Sigma_c \pi$ interaction~\cite{Nieves:2019nol,Lu:2014ina}, whose threshold is located much closer than the mass of the bare three-quark state, or because the ldof in its inner structure are coupled to the unnatural $j_{\rm ldof}^P=0^-$  quantum numbers which gives rise to a double-pole pattern for this resonance analog to that established for the $\Lambda(1405)$~\cite{Garcia-Recio:2008rjt,Romanets:2012hm}. To shed light into this problem, it would require new LQCD simulations giving rise to more accurate determinations of the masses of the $\mcg$ and $\mce$, with precision better than the mass difference with the $\Sigma_c \pi$ and $\Sigma_c^*\pi$ thresholds, and using not only three-quark interpolating operators, but also other ones with  larger overlaps to hadron-molecular degrees of freedom.  

\begin{figure}
\includegraphics[width=1.0\textwidth]{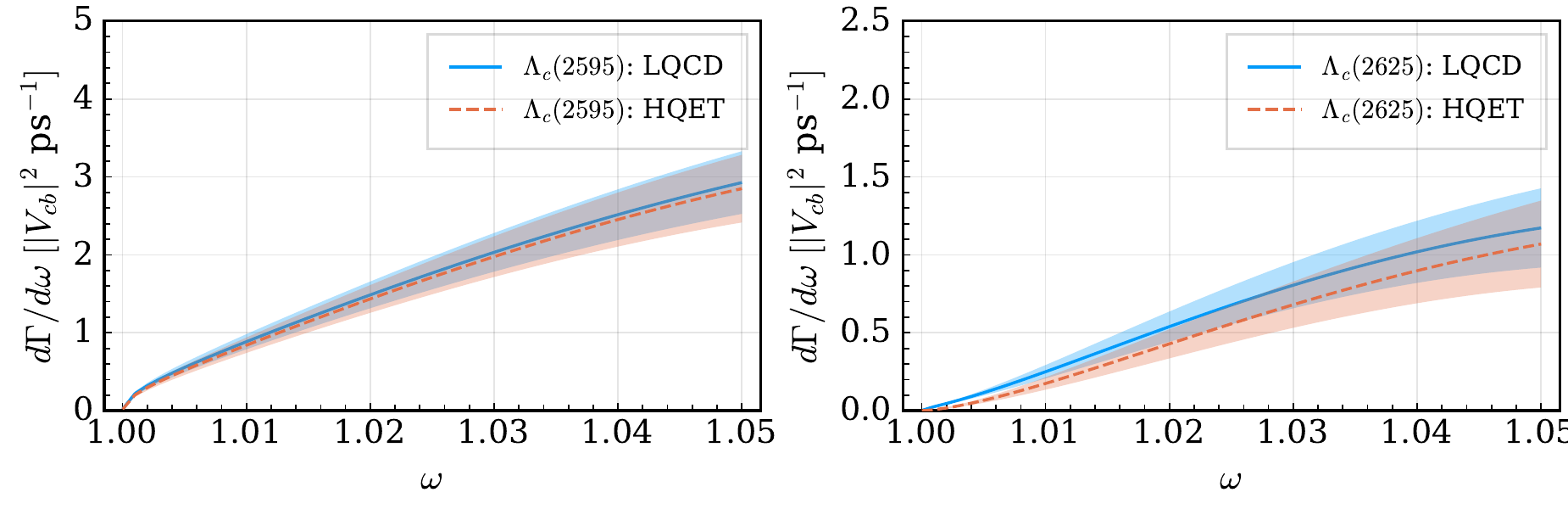}\\
\includegraphics[width=1.0\textwidth]{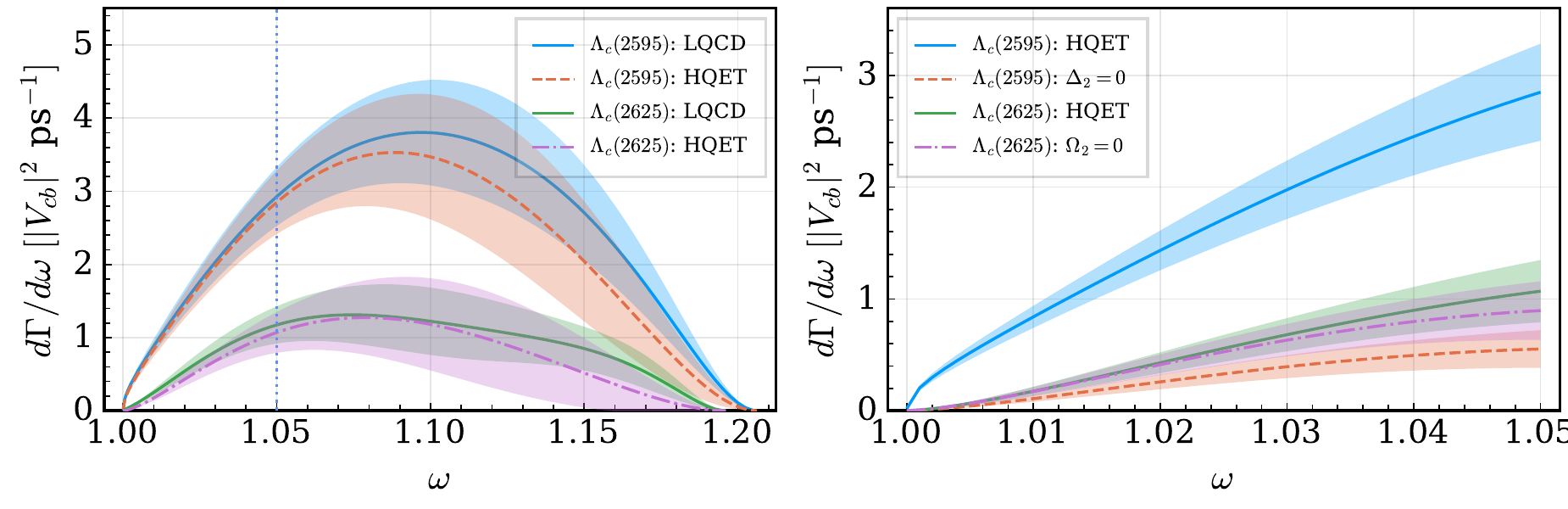}\\
\caption{SM $d\Gamma[\Lambda_b\to \lamcg\tau \bar\nu_\tau]/d\omega$ and $d\Gamma[\Lambda_b\to \lamce\tau \bar\nu_\tau]/d\omega$ differential decay widths, as function of $\omega$. We show distributions obtained using the full set of LQCD form-factors computed in Ref.~\cite{Meinel:2021mdj} (LQCD) and the expressions (HQET) of Eqs.~\eqref{eq:gamma_diff}-\eqref{eq:gamma_diff2}, employing the $\Delta_{1,2}$ and $\Omega_{1,2}$  functions determined in Eq.~\eqref{eq:nDeltaOmega}. For the latter scheme, in the  right-bottom panel, we compare the full $\omc$ HQET results with those obtained after setting to zero the sub-leading $\Delta_2$ and $\Omega_2$ functions. }\label{fig:diff-distr}
\end{figure}

 Coming back to Fig.~\ref{fig:delta_vs_omega}, we observe that both $\Delta_2$ and $\Omega_2$ are also smaller (absolute value) than $\Delta_1 $ and $\Omega_1$, respectively, near zero recoil. Actually, the ratios $\Delta_2/\Delta_1$ and  $\Omega_2/\Omega_1$ (right-bottom plot) have a typical size of order $\omc$, which could be already large for $\omega=1.05$, as we mentioned in the discussion of  Eq.~\eqref{eq:nDeltaOmega}. This supports the compatibility of the LQCD and  $\omc$ HQET predictions for the form-factors, discussed in the previous subsection. Actually in the two top plots and the left-bottom one of Fig.~\ref{fig:diff-distr}, we compare the SM $d\Gamma[\Lambda_b\to \Lambda_{c,J^P}^*\tau \bar\nu_\tau]/d\omega$ distributions obtained using the full set of LQCD form-factors computed in Ref.~\cite{Meinel:2021mdj} and the $\omc$ HQET expressions of Eqs.~\eqref{eq:gamma_diff}-\eqref{eq:gamma_diff2}, employing the $\Delta_{1,2}$ and $\Omega_{1,2}$ functions determined in Eq.~\eqref{eq:nDeltaOmega}. We find a good agreement, not only up to $\w=1.05$ (region of validity of the analysis of Ref.~\cite{Meinel:2021mdj}), but also for the whole phase-space available in the semileptonic decay in the tau mode.  This latter result confirms that the $\Delta_{1,2}$ and $\Omega_{1,2}$ functions, empirically determined from the LQCD form-factors,  contain all type of contributions suppressed by any power of the charm quark mass. Note also that in the left-bottom plot of Fig.~\ref{fig:diff-distr}, some discrepancies between LQCD and HQET predictions can be observed for $\omega> 1.1$, which should be expected, since in that kinematical region, the $\mathcal{O}(|\vec{q}\,|/m_b)$ corrections could start to become sizable. Nevertheless, we should stress that predictions above $\omega=1.05$ might not be correct, because this region is beyond the limit of reliability of the LQCD form-factors obtained in  Refs.~\cite{Meinel:2021mdj,Meinel:2021rbm}.

On the other hand, we can see in Fig.~\ref{fig:delta_vs_omega} that  in the vicinity of zero recoil, $\Delta_2/\Delta_1$ is higher than $\Omega_2/\Omega_1$, reaching the former ratio values around 0.3-0.35. Moreover, since $\Delta_2$  accounts for a S-wave term to  the $\lambb\to \lcg$ decay amplitude (see discussion of Eq.~\eqref{eq:HQratio}), which becomes dominant close to $\w=1$, we see in the right-bottom panel of  Fig.~\ref{fig:diff-distr} that this sub-leading IW function has an enormous numerical impact in the $\Lambda_b\to \lamcg$ differential decay distribution, and its contribution becomes totally dominant.   This result confirms the findings of Refs.~\cite{Leibovich:1997az,Nieves:2019kdh}, obtained using the soliton model derived in Ref.~\cite{Leibovich:1997az}, where the large size of these finite charm quark mass HQSS breaking terms on the $\lamcg$ differential decay distribution was firstly pointed out (see for instance Fig.~2 of Ref.~\cite{Nieves:2019kdh}). In sharp contrast, the effects  on the $\lamce$ distribution of the sub-leading  $\Omega_2$ function are small below $\w=1.05$ (right-bottom panel of Fig.~\ref{fig:diff-distr}).

From the above discussion, we naturally find an explanation  for large deviations of the   $\frac{d\Gamma[\Lambda_b\to \lcg]/d\omega}{d\Gamma[\Lambda_b\to \lce]/d\omega}$ ratio from the value of 1/2  predicted in the $m_Q\to \infty$ limit assuming  that the $\lcg$ and $\lce$ are the two members of the $j_{\rm ldof}^P=1^-$ HQSS doublet. Actually, this ratio at $\w=1.05$ takes values around 2.5 and much larger ones in the near-zero recoil regime. This tension between LQCD data and the HQET predictions triggered  the claim in Refs.~\cite{Papucci:2021pmj,Meinel:2021mdj} of unexpectedly large HQSS-violating terms, with potentially
large $1/m_c^2$ corrections, near zero recoil\footnote{The authors of Refs.~\cite{Papucci:2021pmj,Meinel:2021mdj} pointed out that such large HQSS-violating terms cannot persist uniformly
over the full recoil spectrum,  as they would be incompatible
with the measurement of the ratio of total decay rates in the muon mode~\cite{CDF:2008hqh}
\be
\frac{\Gamma[\lambb \to \lcg \mu \bar\nu_\mu]}{\Gamma[\lambb \to \lce \mu \bar\nu_\mu]} = 0.6 \pm 0.2 ^{+0.5}_{-0.3}
\ee
We note, however, that at the end of the spectrum in the semileptonic muon mode $|\vec{q}\,|\sim 2.2$ GeV and, thus talking about corrections suppressed by the charm quark mass is meaningless.}.
We assign here these huge HQSS breaking corrections, not necessarily just to big values of the $\Delta_2(\w)/\Delta_1(\w)$ ratio of leading to sub-leading IW functions, but also to the fact that the $\Delta_2(\omega)$ form-factor accounts for a S-wave term to the $\lambb\to \lcg$ decay amplitude, while  the $\lambb\to \lce$ semileptonic decay proceeds necessarily in P-wave, and it is therefore suppressed by a factor $(\omega^2-1)^3/2$ at zero recoil\footnote{We remind here that the in the $\lambb$-rest frame the three-momentum of the outgoing hadron ($\Lambda^*_c$) in the semileptonic decay is proportional to $\sqrt{\omega^2-1}$.}. 

 Below $\omega\sim 1.05$, when $\Delta_2$ and $\Omega_2$ are set to zero, we recover within errors the value of 1/2 for the ratio of differential decay widths,  see right-bottom plot of Fig.~\ref{fig:diff-distr}. This is so because, taking into account uncertainties, $\Delta_1(\w)\sim -\Omega_1(\omega)$ even in the presence of $\omc$ corrections to these leading IW functions (see top left panel of Fig.~\ref{fig:delta_vs_omega}). 
This is in accordance with the $\lamcg$ and $\lamce$ being HQSS partners.

\section{Summary}\label{sec:summary}
We have used the $\omc$ HQET scheme derived in Ref.~\cite{Nieves:2019kdh}, in which invariance under rotations of the spin of the  $b$ quark is preserved, to study the  matrix elements of the scalar, pseudoescalar, vector, axial-vector and tensor currents between the $\lambb$ ground state and the  odd parity charm  $\lamcg$ and $\lamce$ resonances. Only $\mathcal{O}(\Lambda_\text{QCD}/m_b)$ and perturbative QCD short distance corrections are neglected. 
There are only four independent functions $\Delta_{1,2}(\w)$ and $\Omega_{1,2}(\w)$ for the $\lambb\to\lcg$ and $\lambb\to\lce$ semileptonic decays, respectively, which are determined from a recent LQCD computation of the corresponding helicity form-factors~\cite{Meinel:2021rbm,Meinel:2021mdj}.   We have shown that in the near-zero recoil regime, this $\omc$ HQET scheme describes reasonably well, taking into account systematic uncertainties,  the  results for the total of 24 form-factors obtained in the LQCD studies of  Refs.~\cite{Meinel:2021rbm,Meinel:2021mdj}. 

We have found that the differences between $\Delta_1$ and $-\Omega_1$ are small compared to any,   $\Delta_1$ or $\Omega_1$, of these leading IW functions, and can be naturally attributed to $\omc$ contributions. This gives some support, or at least does not contradict, to the scenario in which the $\lamcg$ and $\lamce$ resonances  might form the lowest-lying  HQSS $j_{\rm ldof}^P=1^-$ doublet. However, we have argued that the available LQCD description of these two resonances is not accurate enough to disentangle the possible effects of the $\Sigma_c \pi$ and $\Sigma_c^*\pi$ thresholds, located only few MeV above the position of these excited states, especially in the case of the $\lamcg$ and  $\Sigma_c \pi$~\cite{Nieves:2019nol}. The claim made in this latter reference  that the $\lamcg$ and $\lamce$ resonances are not HQSS partners cannot be discarded using the LQCD data studied here. To clarify this problem, new and more precise LQCD simulations capable of elucidating the role played by the $\Sigma_c \pi$ and $\Sigma_c^*\pi$ channels  and that make use, not only of three-quark interpolating fields, but also of other operators with larger overlaps to hadron-molecular degrees of freedom will be required.

Finally, we have naturally given an explanation  to the  large deviations, near zero recoil,  of the   $\frac{d\Gamma[\Lambda_b\to \lcg]/d\omega}{d\Gamma[\Lambda_b\to \lce]/d\omega}$ ratio from 1/2, value predicted in the $m_Q\to \infty$ limit assuming  that the $\lcg$ and $\lce$ are the two members of the $j_{\rm ldof}^P=1^-$ HQSS doublet. We have related these huge HQSS breaking corrections, not necessarily  to just big values of the $\Delta_2(\w)/\Delta_1(\w)$ ratio of  sub-leading to leading IW functions, 
but also to  the S-wave character of the contributions to the $\lambb\to \lcg$ decay amplitude driven by  $\Delta_2(\w)$.

\section*{Acknowledgements}
 This research has been supported  by the Spanish Ministerio de Ciencia e Innovaci\'on (MICINN)
and the European Regional Development Fund (ERDF) under contracts PID2020-112777GB-I00 and PID2019-105439GB-C22, 
the EU STRONG-2020 project under the program H2020-INFRAIA-2018-1, 
grant agreement no. 824093 and by  Generalitat Valenciana under contract PROMETEO/2020/023.

\newpage 

\appendix

\section{LQCD form-factors}\label{app:LQCDFF}

The relations between the form factors in Eqs.~\eqref{eq:formfactor12} and \eqref{eq:formfactor32} and the LQCD ones computed in Ref.~\cite{Meinel:2021mdj} are for $\lambb\to\lamcg$,
\bea\label{eq:htod}
d_S \al = \al \fzero\frac{\mb+\mcg}{m_b-m_c},\nno 
d_P \al = \al \gzero\frac{\mb-\mcg}{m_b+m_c},\nno
d_{V_1}\al = \al -\fperp, \nno
d_{V_2}\al = \al \mb \left[ \frac{\mb+\mcg}{q^2} \fzero + \frac{\mb-\mcg}{s_-}\left( 1-\frac{\mb^2-\mcg^2}{q^2}\right)\fplus + \frac{2\mcg}{s_-}\fperp \right] , \nno 
d_{V_3} \al = \al \mcg\left[ -\frac{\mb+\mcg}{q^2}\fzero + \frac{\mb-\mcg}{s_-}\left(1+\frac{\mb^2-\mcg^2}{q^2}\right)\fplus-\frac{2\mb}{s_-}\fperp \right], \nno 
d_{A_1} \al = \al - \gperp, \nno
d_{A_2} \al = \al -\mb\left[ \frac{\mb-\mcg}{q^2}\gzero +\frac{\mb+\mcg}{s_+}\left( 1-\frac{\mb^2-\mcg^2}{q^2}\right)\gplus -\frac{2\mcg}{s_+} \gperp \right], \nno
d_{A_3} \al = \al \mcg\left[ \frac{\mb-\mcg}{q^2} \gzero - \frac{\mb+\mcg}{s_+}\left(1+\frac{\mb^2-\mcg^2}{q^2}\right)\gplus+\frac{2\mb}{s_+}\gperp \right], \nno 
d_{T_1} \al = \al 2\mb^2\left( \frac{1}{s_-}h_+^{(\frac12^-)}-\frac{(\mb-\mcg)^2}{q^2 s_-}h_\perp^{(\frac12^-)}-\frac{1}{s_+}\hh_+^{(\frac12^-)}+\frac{(\mb+\mcg)^2}{q^2 s_+}\hh_\perp^{(\frac12^-)}\right),\nno 
d_{T_2} \al = \al 2\mb\left( \frac{\mcg}{s_-}h_+^{(\frac12^-)} -\frac{(\mb-\mcg)(\mb^2-\mcg^2-q^2)}{2q^2s_-}h_\perp^{(\frac12^-)} +\frac{\mb+\mcg}{2q^2}\hh_\perp^{(\frac12^-)}\right),\nno
d_{T_3}\al = \al 2\mb\left( -\frac{\mb}{s_-}h_+^{(\frac12^-)} +\frac{(\mb-\mcg)(\mb^2-\mcg^2+q^2)}{2q^2s_-}h_\perp^{(\frac12^-)} -\frac{\mb+\mcg}{2q^2}\hh_\perp^{(\frac12^-)}\right),\nno
d_{T_4}\al = \al h_+^{(\frac12^-)},
\eea
and for $\lambb\to\lamce$,
\bea\label{eq:htol}
l_S \al = \al f_0^{(\frac32^-)}\frac{\mb\mce}{s_+}\frac{\mb-\mce}{m_b-m_c}, \nno
l_P \al = \al g_0^{(\frac32^-)}\frac{\mb\mce}{s_-}\frac{\mb+\mce}{m_b+m_c}, \nno
l_{V_1} \al = \al \frac{\mb\mce}{s_-} (f_\perp^{(\frac32^-)}+f_{\perp^\prime}^{(\frac32^-)}), \nno
l_{V_2} \al = \al \frac{\mb^2\mce}{s_+ s_-}\bigg[ f_0^{(\frac32^-)}\frac{(\mb-\mce)s_-}{q^2}+ f_+^{(\frac32^-)}(\mb+\mce)\left(1-\frac{\mb^2-\mce^2}{q^2}\right)\nno
\al \al -2\mce f_\perp^{(\frac32^-)}+2\mce f_{\perp^\prime}^{(\frac32^-)}\bigg], \nno 
l_{V_3} \al = \al \frac{\mb\mce^2}{s_+ s_-}\bigg[ -f_0^{(\frac32^-)} \frac{(\mb-\mce)s_-}{q^2}+f_+^{(\frac32^-)}(\mb+\mce)\left(1+\frac{\mb^2-\mce^2}{q^2}\right) \nno 
\al \al -2\mb f_\perp^{(\frac32^-)}+2(\mb-\frac{s_+}{\mce})f_{\perp^\prime}^{(\frac32^-)} \bigg], \nno
l_{V_4} \al = \al f_{\perp^\prime}^{(\frac32^-)}, \nno
l_{A_1} \al = \al \frac{\mb\mce}{s_+ }(g_\perp^{(\frac32^-)}+g_{\perp^\prime}^{(\frac32^-)}), \nno
l_{A_2} \al = \al \frac{\mb^2\mce}{s_+s_-}\bigg[ -\frac{(\mb+\mce)s_+}{q^2}g_0^{(\frac32^-)}+\frac{(\mb-\mce)(\mb^2-\mce^2-q^2)}{q^2}g_+^{(\frac32^-)} \nno 
\al \al - 2\mce g_\perp^{(\frac32^-)} +2\mce g_{\perp^\prime}^{(\frac32^-)} \bigg], \nno
l_{A_3} \al = \al \frac{\mb \mce^2}{s_+s_-}\bigg[ \frac{(\mb+\mce)s_+}{q^2}g_0^{(\frac32^-)} -\frac{(\mb-\mce)(\mb^2-\mce^2+q^2)}{q^2}g_+^{(\frac32^-)} \nno 
\al \al + 2\mb g_\perp^{(\frac32^-)} -2(\mb +\frac{s_-}{\mce})g_{\perp^\prime}^{(\frac32^-)}\bigg], \nno
l_{A_4}  \al = \al g_{\perp^\prime}^{(\frac32^-)}, \nno
l_{T_1}\al = \al \frac{2\mb^3}{q^2s_-s_+}\bigg( \mce \left[ (\mb+\mce)^2 h_\perp^{(\frac32^-)} + q^2(\hh_+^{(\frac32^-)}-h_+^{(\frac32^-)})- (\mb-\mce)^2 \hh_\perp^{(\frac32^-)}\right] \nno 
\al \al + h_{\perp^\prime}^{(\frac32^-)}(\mb+\mce)(\mb^2+\mb\mce-q^2) +(\mb-\mce)(\mb^2-\mb\mce-q^2)\hh_{\perp^\prime}^{(\frac32^-)}\bigg), \nno
l_{T_2} \al = \al \frac{\mb^2\mce}{q^2 s_-s_+}\bigg( s_+(\mb+\mce)(h_\perp^{(\frac32^-)} +h_{\perp^\prime}^{(\frac32^-)})  \nno 
\al \al +(\mb-\mce)(\mb^2-\mce^2-q^2)(\hh_{\perp^\prime}^{(\frac32^-)}-\hh_\perp^{(\frac32^-)}) +2\mce q^2\hh_+^{(\frac32^-)}\bigg), \\
l_{T_3} \al = \al -\frac{\mb}{q^2s_-s_+}\bigg( s_+\left[\mb^3-q^2(\mb+\mce)+\mce^3\right] h_{\perp^\prime}^{(\frac32^-)}  \nno 
\al \al + \mb\mce\left[ s_+ (\mb+\mce)h_\perp^{(\frac32^-)}-(\mb-\mce)(\mb^2-\mce^2+q^2)\hh_{\perp}^{(\frac32^-)} + 2\mb q^2\hh_+^{(\frac32^-)} \right] \nno 
\al \al + (\mb-\mce)(\mb^2-\mb\mce+\mce^2-q^2)(\mb^2-\mce^2+q^2)\hh_{\perp^\prime}^{(\frac32^-)}\bigg),\nno
l_{T_4}\al = \al -\frac{\mce}{q^2s_+}\bigg( s_+(\mb+\mce)h_{\perp^\prime}^{(\frac32^-)} +(\mb-\mce)(\mb^2-\mce^2+q^2)\hh_{\perp^\prime}^{(\frac32^-)} -\mb q^2\hh_+^{(\frac32^-)}\bigg)  \nno
l_{T_5} \al = \al -\frac{s_+ h_{\perp^\prime}^{(\frac32^-)}(\mb+\mce)+(\mb-\mce)(\mb^2-\mce^2+q^2)\hh_{\perp^\prime}^{(\frac32^-)}}{2\mb q^2} ,\nno 
l_{T_6} \al = \al \frac{h_{\perp^\prime}^{(\frac32^-)}(\mb+\mce)^2 + (\mb-\mce)^2 \hh_{\perp^\prime}^{(\frac32^-)}}{q^2},
\eea
where $q=p-p^\prime$, $s_\pm = (\mb \pm m_{\Lambda_c^*})^2-q^2$ and $m_b$($m_c$) the mass of $b$($c$) quark. The equations of motion of the heavy quarks ($b$ and $c$ quark) have been used to derive the scalar and pseudoscalar form factors.

\section{HQET predictions for the form factors}\label{app:ff:hqss}

The non-vanishing form factors in HQET up to $\omc$ read
\bea\label{eq:ff:HQET}
\al d_{S} = -\dfrac{1}{\sqrt{3}}[(1+\w) \Delta_1 +\Delta_2] , \qquad \al d_P = -\frac{1}{\sqrt{3}}[(\w-1)\Delta_1 +\Delta_2], \nno
\al d_{V_1} = \dfrac{1}{\sqrt{3}}[(\w-1)\Delta_1 +\Delta_2 ] , \qquad \al d_{V_2} = -\frac{2}{\sqrt{3}}\Delta_1, \nno 
\al d_{A_1} = \dfrac{1}{\sqrt{3}}[(\w+1)\Delta_1 + \Delta_2 ], \qquad \al d_{A_2} = -\frac{2}{\sqrt{3}}\Delta_1, \nno 
\al d_{T_2} = {-}\dfrac{2}{\sqrt{3}} \Delta_1 , \qquad \qquad\qquad \quad \al d_{T_4}  = -\frac{1}{\sqrt{3}}[(\w-1)\Delta_1 + \Delta_2 ], \nno
\al l_S = \Omega_1 + (\w-1)\Omega_2, \qquad\qquad \al l_P = \Omega_1 + (\w+1)\Omega_2, \nno
\al l_{V_1} = \Omega_1 + (\w+1)\Omega_2, \qquad\qquad \al l_{V_2} = -2\Omega_2 , \nno 
\al l_{A_1} = \Omega_1 + (\w-1)\Omega_2, \qquad\qquad \al l_{A_2} = -2\Omega_2 , \nno 
\al l_{T_2} = 2\Omega_2 ,\qquad \qquad\qquad\quad \qquad \al l_{T_4} = \Omega_1 + (\w-1)\Omega_2.
\eea
The form factors not listed above, i.e. $d_{V_3}$, $d_{A_3}$, $d_{T_1}$, $d_{T_3}$, $l_{V_3}$, $l_{V_4}$, $l_{A_3}$, $l_{A_4}$, $l_{T_1}$, $l_{T_3}$, $l_{T_5}$, $l_{T_6}$, vanish at this order.

\newpage

\bibliography{hqss.bib}

\begin{thebibliography}{46}%
\makeatletter
\providecommand \@ifxundefined [1]{%
 \@ifx{#1\undefined}
}%
\providecommand \@ifnum [1]{%
 \ifnum #1\expandafter \@firstoftwo
 \else \expandafter \@secondoftwo
 \fi
}%
\providecommand \@ifx [1]{%
 \ifx #1\expandafter \@firstoftwo
 \else \expandafter \@secondoftwo
 \fi
}%
\providecommand \natexlab [1]{#1}%
\providecommand \enquote  [1]{``#1''}%
\providecommand \bibnamefont  [1]{#1}%
\providecommand \bibfnamefont [1]{#1}%
\providecommand \citenamefont [1]{#1}%
\providecommand \href@noop [0]{\@secondoftwo}%
\providecommand \href [0]{\begingroup \@sanitize@url \@href}%
\providecommand \@href[1]{\@@startlink{#1}\@@href}%
\providecommand \@@href[1]{\endgroup#1\@@endlink}%
\providecommand \@sanitize@url [0]{\catcode `\\12\catcode `\$12\catcode
  `\&12\catcode `\#12\catcode `\^12\catcode `\_12\catcode `\%12\relax}%
\providecommand \@@startlink[1]{}%
\providecommand \@@endlink[0]{}%
\providecommand \url  [0]{\begingroup\@sanitize@url \@url }%
\providecommand \@url [1]{\endgroup\@href {#1}{\urlprefix }}%
\providecommand \urlprefix  [0]{URL }%
\providecommand \Eprint [0]{\href }%
\providecommand \doibase [0]{https://doi.org/}%
\providecommand \selectlanguage [0]{\@gobble}%
\providecommand \bibinfo  [0]{\@secondoftwo}%
\providecommand \bibfield  [0]{\@secondoftwo}%
\providecommand \translation [1]{[#1]}%
\providecommand \BibitemOpen [0]{}%
\providecommand \bibitemStop [0]{}%
\providecommand \bibitemNoStop [0]{.\EOS\space}%
\providecommand \EOS [0]{\spacefactor3000\relax}%
\providecommand \BibitemShut  [1]{\csname bibitem#1\endcsname}%
\let\auto@bib@innerbib\@empty
\bibitem [{\citenamefont {Leibovich}\ and\ \citenamefont
  {Stewart}(1998)}]{Leibovich:1997az}%
  \BibitemOpen
  \bibfield  {author} {\bibinfo {author} {\bibfnamefont {A.~K.}\ \bibnamefont
  {Leibovich}}\ and\ \bibinfo {author} {\bibfnamefont {I.~W.}\ \bibnamefont
  {Stewart}},\ }\bibfield  {title} {\bibinfo {title} {{Semileptonic Lambda(b)
  decay to excited Lambda(c) baryons at order Lambda(QCD) / m(Q)}},\ }\href
  {https://doi.org/10.1103/PhysRevD.57.5620} {\bibfield  {journal} {\bibinfo
  {journal} {Phys. Rev. D}\ }\textbf {\bibinfo {volume} {57}},\ \bibinfo
  {pages} {5620} (\bibinfo {year} {1998})},\ \Eprint
  {https://arxiv.org/abs/hep-ph/9711257} {arXiv:hep-ph/9711257} \BibitemShut
  {NoStop}%
\bibitem [{\citenamefont {Pervin}\ \emph {et~al.}(2005)\citenamefont {Pervin},
  \citenamefont {Roberts},\ and\ \citenamefont {Capstick}}]{Pervin:2005ve}%
  \BibitemOpen
  \bibfield  {author} {\bibinfo {author} {\bibfnamefont {M.}~\bibnamefont
  {Pervin}}, \bibinfo {author} {\bibfnamefont {W.}~\bibnamefont {Roberts}},\
  and\ \bibinfo {author} {\bibfnamefont {S.}~\bibnamefont {Capstick}},\
  }\bibfield  {title} {\bibinfo {title} {{Semileptonic decays of heavy lambda
  baryons in a quark model}},\ }\href
  {https://doi.org/10.1103/PhysRevC.72.035201} {\bibfield  {journal} {\bibinfo
  {journal} {Phys. Rev. C}\ }\textbf {\bibinfo {volume} {72}},\ \bibinfo
  {pages} {035201} (\bibinfo {year} {2005})},\ \Eprint
  {https://arxiv.org/abs/nucl-th/0503030} {arXiv:nucl-th/0503030} \BibitemShut
  {NoStop}%
\bibitem [{\citenamefont {Yoshida}\ \emph {et~al.}(2015)\citenamefont
  {Yoshida}, \citenamefont {Hiyama}, \citenamefont {Hosaka}, \citenamefont
  {Oka},\ and\ \citenamefont {Sadato}}]{Yoshida:2015tia}%
  \BibitemOpen
  \bibfield  {author} {\bibinfo {author} {\bibfnamefont {T.}~\bibnamefont
  {Yoshida}}, \bibinfo {author} {\bibfnamefont {E.}~\bibnamefont {Hiyama}},
  \bibinfo {author} {\bibfnamefont {A.}~\bibnamefont {Hosaka}}, \bibinfo
  {author} {\bibfnamefont {M.}~\bibnamefont {Oka}},\ and\ \bibinfo {author}
  {\bibfnamefont {K.}~\bibnamefont {Sadato}},\ }\bibfield  {title} {\bibinfo
  {title} {{Spectrum of heavy baryons in the quark model}},\ }\href
  {https://doi.org/10.1103/PhysRevD.92.114029} {\bibfield  {journal} {\bibinfo
  {journal} {Phys. Rev. D}\ }\textbf {\bibinfo {volume} {92}},\ \bibinfo
  {pages} {114029} (\bibinfo {year} {2015})},\ \Eprint
  {https://arxiv.org/abs/1510.01067} {arXiv:1510.01067 [hep-ph]} \BibitemShut
  {NoStop}%
\bibitem [{\citenamefont {Nagahiro}\ \emph {et~al.}(2017)\citenamefont
  {Nagahiro}, \citenamefont {Yasui}, \citenamefont {Hosaka}, \citenamefont
  {Oka},\ and\ \citenamefont {Noumi}}]{Nagahiro:2016nsx}%
  \BibitemOpen
  \bibfield  {author} {\bibinfo {author} {\bibfnamefont {H.}~\bibnamefont
  {Nagahiro}}, \bibinfo {author} {\bibfnamefont {S.}~\bibnamefont {Yasui}},
  \bibinfo {author} {\bibfnamefont {A.}~\bibnamefont {Hosaka}}, \bibinfo
  {author} {\bibfnamefont {M.}~\bibnamefont {Oka}},\ and\ \bibinfo {author}
  {\bibfnamefont {H.}~\bibnamefont {Noumi}},\ }\bibfield  {title} {\bibinfo
  {title} {{Structure of charmed baryons studied by pionic decays}},\ }\href
  {https://doi.org/10.1103/PhysRevD.95.014023} {\bibfield  {journal} {\bibinfo
  {journal} {Phys. Rev. D}\ }\textbf {\bibinfo {volume} {95}},\ \bibinfo
  {pages} {014023} (\bibinfo {year} {2017})},\ \Eprint
  {https://arxiv.org/abs/1609.01085} {arXiv:1609.01085 [hep-ph]} \BibitemShut
  {NoStop}%
\bibitem [{\citenamefont {B\"oer}\ \emph {et~al.}(2018)\citenamefont {B\"oer},
  \citenamefont {Bordone}, \citenamefont {Graverini}, \citenamefont {Owen},
  \citenamefont {Rotondo},\ and\ \citenamefont {Van~Dyk}}]{Boer:2018vpx}%
  \BibitemOpen
  \bibfield  {author} {\bibinfo {author} {\bibfnamefont {P.}~\bibnamefont
  {B\"oer}}, \bibinfo {author} {\bibfnamefont {M.}~\bibnamefont {Bordone}},
  \bibinfo {author} {\bibfnamefont {E.}~\bibnamefont {Graverini}}, \bibinfo
  {author} {\bibfnamefont {P.}~\bibnamefont {Owen}}, \bibinfo {author}
  {\bibfnamefont {M.}~\bibnamefont {Rotondo}},\ and\ \bibinfo {author}
  {\bibfnamefont {D.}~\bibnamefont {Van~Dyk}},\ }\bibfield  {title} {\bibinfo
  {title} {{Testing lepton flavour universality in semileptonic $\Lambda_b \to
  \Lambda_c^*$ decays}},\ }\href {https://doi.org/10.1007/JHEP06(2018)155}
  {\bibfield  {journal} {\bibinfo  {journal} {JHEP}\ }\textbf {\bibinfo
  {volume} {06}},\ \bibinfo {pages} {155}},\ \Eprint
  {https://arxiv.org/abs/1801.08367} {arXiv:1801.08367 [hep-ph]} \BibitemShut
  {NoStop}%
\bibitem [{\citenamefont {Meinel}\ and\ \citenamefont
  {Rendon}(2021)}]{Meinel:2021rbm}%
  \BibitemOpen
  \bibfield  {author} {\bibinfo {author} {\bibfnamefont {S.}~\bibnamefont
  {Meinel}}\ and\ \bibinfo {author} {\bibfnamefont {G.}~\bibnamefont
  {Rendon}},\ }\bibfield  {title} {\bibinfo {title} {{$\Lambda_b \to
  \Lambda_c^*(2595,2625)\ell^-\bar{\nu}$form factors from lattice QCD}},\
  }\href {https://doi.org/10.1103/PhysRevD.103.094516} {\bibfield  {journal}
  {\bibinfo  {journal} {Phys. Rev. D}\ }\textbf {\bibinfo {volume} {103}},\
  \bibinfo {pages} {094516} (\bibinfo {year} {2021})},\ \Eprint
  {https://arxiv.org/abs/2103.08775} {arXiv:2103.08775 [hep-lat]} \BibitemShut
  {NoStop}%
\bibitem [{\citenamefont {Papucci}\ and\ \citenamefont
  {Robinson}(2022)}]{Papucci:2021pmj}%
  \BibitemOpen
  \bibfield  {author} {\bibinfo {author} {\bibfnamefont {M.}~\bibnamefont
  {Papucci}}\ and\ \bibinfo {author} {\bibfnamefont {D.~J.}\ \bibnamefont
  {Robinson}},\ }\bibfield  {title} {\bibinfo {title} {{Form factor counting
  and HQET matching for new physics in
  \ensuremath{\Lambda}b\textrightarrow{}\ensuremath{\Lambda}c*l\ensuremath{\nu}}},\
  }\href {https://doi.org/10.1103/PhysRevD.105.016027} {\bibfield  {journal}
  {\bibinfo  {journal} {Phys. Rev. D}\ }\textbf {\bibinfo {volume} {105}},\
  \bibinfo {pages} {016027} (\bibinfo {year} {2022})},\ \Eprint
  {https://arxiv.org/abs/2105.09330} {arXiv:2105.09330 [hep-ph]} \BibitemShut
  {NoStop}%
\bibitem [{\citenamefont {Migura}\ \emph {et~al.}(2006)\citenamefont {Migura},
  \citenamefont {Merten}, \citenamefont {Metsch},\ and\ \citenamefont
  {Petry}}]{Migura:2006ep}%
  \BibitemOpen
  \bibfield  {author} {\bibinfo {author} {\bibfnamefont {S.}~\bibnamefont
  {Migura}}, \bibinfo {author} {\bibfnamefont {D.}~\bibnamefont {Merten}},
  \bibinfo {author} {\bibfnamefont {B.}~\bibnamefont {Metsch}},\ and\ \bibinfo
  {author} {\bibfnamefont {H.-R.}\ \bibnamefont {Petry}},\ }\bibfield  {title}
  {\bibinfo {title} {{Charmed baryons in a relativistic quark model}},\ }\href
  {https://doi.org/10.1140/epja/i2006-10017-9} {\bibfield  {journal} {\bibinfo
  {journal} {Eur. Phys. J. A}\ }\textbf {\bibinfo {volume} {28}},\ \bibinfo
  {pages} {41} (\bibinfo {year} {2006})},\ \Eprint
  {https://arxiv.org/abs/hep-ph/0602153} {arXiv:hep-ph/0602153} \BibitemShut
  {NoStop}%
\bibitem [{\citenamefont {Garcilazo}\ \emph {et~al.}(2007)\citenamefont
  {Garcilazo}, \citenamefont {Vijande},\ and\ \citenamefont
  {Valcarce}}]{Garcilazo:2007eh}%
  \BibitemOpen
  \bibfield  {author} {\bibinfo {author} {\bibfnamefont {H.}~\bibnamefont
  {Garcilazo}}, \bibinfo {author} {\bibfnamefont {J.}~\bibnamefont {Vijande}},\
  and\ \bibinfo {author} {\bibfnamefont {A.}~\bibnamefont {Valcarce}},\
  }\bibfield  {title} {\bibinfo {title} {{Faddeev study of heavy baryon
  spectroscopy}},\ }\href {https://doi.org/10.1088/0954-3899/34/5/014}
  {\bibfield  {journal} {\bibinfo  {journal} {J. Phys. G}\ }\textbf {\bibinfo
  {volume} {34}},\ \bibinfo {pages} {961} (\bibinfo {year} {2007})},\ \Eprint
  {https://arxiv.org/abs/hep-ph/0703257} {arXiv:hep-ph/0703257} \BibitemShut
  {NoStop}%
\bibitem [{\citenamefont {Roberts}\ and\ \citenamefont
  {Pervin}(2008)}]{Roberts:2007ni}%
  \BibitemOpen
  \bibfield  {author} {\bibinfo {author} {\bibfnamefont {W.}~\bibnamefont
  {Roberts}}\ and\ \bibinfo {author} {\bibfnamefont {M.}~\bibnamefont
  {Pervin}},\ }\bibfield  {title} {\bibinfo {title} {{Heavy baryons in a quark
  model}},\ }\href {https://doi.org/10.1142/S0217751X08041219} {\bibfield
  {journal} {\bibinfo  {journal} {Int. J. Mod. Phys. A}\ }\textbf {\bibinfo
  {volume} {23}},\ \bibinfo {pages} {2817} (\bibinfo {year} {2008})},\ \Eprint
  {https://arxiv.org/abs/0711.2492} {arXiv:0711.2492 [nucl-th]} \BibitemShut
  {NoStop}%
\bibitem [{\citenamefont {Arifi}\ \emph {et~al.}(2017)\citenamefont {Arifi},
  \citenamefont {Nagahiro},\ and\ \citenamefont {Hosaka}}]{Arifi:2017sac}%
  \BibitemOpen
  \bibfield  {author} {\bibinfo {author} {\bibfnamefont {A.~J.}\ \bibnamefont
  {Arifi}}, \bibinfo {author} {\bibfnamefont {H.}~\bibnamefont {Nagahiro}},\
  and\ \bibinfo {author} {\bibfnamefont {A.}~\bibnamefont {Hosaka}},\
  }\bibfield  {title} {\bibinfo {title} {{Three-Body Decay of $\Lambda_c^{*}
  (2595)$ and $\Lambda_c^{*} (2625)$ with consideration of $\Sigma_c(2455)\pi$
  and $\Sigma_c^*(2520)\pi$ in intermediate States}},\ }\href
  {https://doi.org/10.1103/PhysRevD.95.114018} {\bibfield  {journal} {\bibinfo
  {journal} {Phys. Rev. D}\ }\textbf {\bibinfo {volume} {95}},\ \bibinfo
  {pages} {114018} (\bibinfo {year} {2017})},\ \Eprint
  {https://arxiv.org/abs/1704.00464} {arXiv:1704.00464 [hep-ph]} \BibitemShut
  {NoStop}%
\bibitem [{\citenamefont {Lutz}\ and\ \citenamefont
  {Kolomeitsev}(2004)}]{Lutz:2003jw}%
  \BibitemOpen
  \bibfield  {author} {\bibinfo {author} {\bibfnamefont {M.~F.~M.}\
  \bibnamefont {Lutz}}\ and\ \bibinfo {author} {\bibfnamefont {E.~E.}\
  \bibnamefont {Kolomeitsev}},\ }\bibfield  {title} {\bibinfo {title} {{On
  charm baryon resonances and chiral symmetry}},\ }\href
  {https://doi.org/10.1016/j.nuclphysa.2003.10.012} {\bibfield  {journal}
  {\bibinfo  {journal} {Nucl. Phys. A}\ }\textbf {\bibinfo {volume} {730}},\
  \bibinfo {pages} {110} (\bibinfo {year} {2004})},\ \Eprint
  {https://arxiv.org/abs/hep-ph/0307233} {arXiv:hep-ph/0307233} \BibitemShut
  {NoStop}%
\bibitem [{\citenamefont {Tolos}\ \emph {et~al.}(2004)\citenamefont {Tolos},
  \citenamefont {Schaffner-Bielich},\ and\ \citenamefont
  {Mishra}}]{Tolos:2004yg}%
  \BibitemOpen
  \bibfield  {author} {\bibinfo {author} {\bibfnamefont {L.}~\bibnamefont
  {Tolos}}, \bibinfo {author} {\bibfnamefont {J.}~\bibnamefont
  {Schaffner-Bielich}},\ and\ \bibinfo {author} {\bibfnamefont
  {A.}~\bibnamefont {Mishra}},\ }\bibfield  {title} {\bibinfo {title}
  {{Properties of D-mesons in nuclear matter within a self-consistent
  coupled-channel approach}},\ }\href
  {https://doi.org/10.1103/PhysRevC.70.025203} {\bibfield  {journal} {\bibinfo
  {journal} {Phys. Rev. C}\ }\textbf {\bibinfo {volume} {70}},\ \bibinfo
  {pages} {025203} (\bibinfo {year} {2004})},\ \Eprint
  {https://arxiv.org/abs/nucl-th/0404064} {arXiv:nucl-th/0404064} \BibitemShut
  {NoStop}%
\bibitem [{\citenamefont {Hofmann}\ and\ \citenamefont
  {Lutz}(2005)}]{Hofmann:2005sw}%
  \BibitemOpen
  \bibfield  {author} {\bibinfo {author} {\bibfnamefont {J.}~\bibnamefont
  {Hofmann}}\ and\ \bibinfo {author} {\bibfnamefont {M.~F.~M.}\ \bibnamefont
  {Lutz}},\ }\bibfield  {title} {\bibinfo {title} {{Coupled-channel study of
  crypto-exotic baryons with charm}},\ }\href
  {https://doi.org/10.1016/j.nuclphysa.2005.08.022} {\bibfield  {journal}
  {\bibinfo  {journal} {Nucl. Phys. A}\ }\textbf {\bibinfo {volume} {763}},\
  \bibinfo {pages} {90} (\bibinfo {year} {2005})},\ \Eprint
  {https://arxiv.org/abs/hep-ph/0507071} {arXiv:hep-ph/0507071} \BibitemShut
  {NoStop}%
\bibitem [{\citenamefont {Mizutani}\ and\ \citenamefont
  {Ramos}(2006)}]{Mizutani:2006vq}%
  \BibitemOpen
  \bibfield  {author} {\bibinfo {author} {\bibfnamefont {T.}~\bibnamefont
  {Mizutani}}\ and\ \bibinfo {author} {\bibfnamefont {A.}~\bibnamefont
  {Ramos}},\ }\bibfield  {title} {\bibinfo {title} {{D mesons in nuclear
  matter: A DN coupled-channel equations approach}},\ }\href
  {https://doi.org/10.1103/PhysRevC.74.065201} {\bibfield  {journal} {\bibinfo
  {journal} {Phys. Rev. C}\ }\textbf {\bibinfo {volume} {74}},\ \bibinfo
  {pages} {065201} (\bibinfo {year} {2006})},\ \Eprint
  {https://arxiv.org/abs/hep-ph/0607257} {arXiv:hep-ph/0607257} \BibitemShut
  {NoStop}%
\bibitem [{\citenamefont {Hofmann}\ and\ \citenamefont
  {Lutz}(2006)}]{Hofmann:2006qx}%
  \BibitemOpen
  \bibfield  {author} {\bibinfo {author} {\bibfnamefont {J.}~\bibnamefont
  {Hofmann}}\ and\ \bibinfo {author} {\bibfnamefont {M.~F.~M.}\ \bibnamefont
  {Lutz}},\ }\bibfield  {title} {\bibinfo {title} {{D-wave baryon resonances
  with charm from coupled-channel dynamics}},\ }\href
  {https://doi.org/10.1016/j.nuclphysa.2006.07.004} {\bibfield  {journal}
  {\bibinfo  {journal} {Nucl. Phys. A}\ }\textbf {\bibinfo {volume} {776}},\
  \bibinfo {pages} {17} (\bibinfo {year} {2006})},\ \Eprint
  {https://arxiv.org/abs/hep-ph/0601249} {arXiv:hep-ph/0601249} \BibitemShut
  {NoStop}%
\bibitem [{\citenamefont {Garcia-Recio}\ \emph {et~al.}(2009)\citenamefont
  {Garcia-Recio}, \citenamefont {Magas}, \citenamefont {Mizutani},
  \citenamefont {Nieves}, \citenamefont {Ramos}, \citenamefont {Salcedo},\ and\
  \citenamefont {Tolos}}]{Garcia-Recio:2008rjt}%
  \BibitemOpen
  \bibfield  {author} {\bibinfo {author} {\bibfnamefont {C.}~\bibnamefont
  {Garcia-Recio}}, \bibinfo {author} {\bibfnamefont {V.~K.}\ \bibnamefont
  {Magas}}, \bibinfo {author} {\bibfnamefont {T.}~\bibnamefont {Mizutani}},
  \bibinfo {author} {\bibfnamefont {J.}~\bibnamefont {Nieves}}, \bibinfo
  {author} {\bibfnamefont {A.}~\bibnamefont {Ramos}}, \bibinfo {author}
  {\bibfnamefont {L.~L.}\ \bibnamefont {Salcedo}},\ and\ \bibinfo {author}
  {\bibfnamefont {L.}~\bibnamefont {Tolos}},\ }\bibfield  {title} {\bibinfo
  {title} {{The s-wave charmed baryon resonances from a coupled-channel
  approach with heavy quark symmetry}},\ }\href
  {https://doi.org/10.1103/PhysRevD.79.054004} {\bibfield  {journal} {\bibinfo
  {journal} {Phys. Rev. D}\ }\textbf {\bibinfo {volume} {79}},\ \bibinfo
  {pages} {054004} (\bibinfo {year} {2009})},\ \Eprint
  {https://arxiv.org/abs/0807.2969} {arXiv:0807.2969 [hep-ph]} \BibitemShut
  {NoStop}%
\bibitem [{\citenamefont {Romanets}\ \emph {et~al.}(2012)\citenamefont
  {Romanets}, \citenamefont {Tolos}, \citenamefont {Garcia-Recio},
  \citenamefont {Nieves}, \citenamefont {Salcedo},\ and\ \citenamefont
  {Timmermans}}]{Romanets:2012hm}%
  \BibitemOpen
  \bibfield  {author} {\bibinfo {author} {\bibfnamefont {O.}~\bibnamefont
  {Romanets}}, \bibinfo {author} {\bibfnamefont {L.}~\bibnamefont {Tolos}},
  \bibinfo {author} {\bibfnamefont {C.}~\bibnamefont {Garcia-Recio}}, \bibinfo
  {author} {\bibfnamefont {J.}~\bibnamefont {Nieves}}, \bibinfo {author}
  {\bibfnamefont {L.~L.}\ \bibnamefont {Salcedo}},\ and\ \bibinfo {author}
  {\bibfnamefont {R.~G.~E.}\ \bibnamefont {Timmermans}},\ }\bibfield  {title}
  {\bibinfo {title} {{Charmed and strange baryon resonances with heavy-quark
  spin symmetry}},\ }\href {https://doi.org/10.1103/PhysRevD.85.114032}
  {\bibfield  {journal} {\bibinfo  {journal} {Phys. Rev. D}\ }\textbf {\bibinfo
  {volume} {85}},\ \bibinfo {pages} {114032} (\bibinfo {year} {2012})},\
  \Eprint {https://arxiv.org/abs/1202.2239} {arXiv:1202.2239 [hep-ph]}
  \BibitemShut {NoStop}%
\bibitem [{\citenamefont {Lu}\ \emph {et~al.}(2015)\citenamefont {Lu},
  \citenamefont {Zhou}, \citenamefont {Chen}, \citenamefont {Xie},\ and\
  \citenamefont {Geng}}]{Lu:2014ina}%
  \BibitemOpen
  \bibfield  {author} {\bibinfo {author} {\bibfnamefont {J.-X.}\ \bibnamefont
  {Lu}}, \bibinfo {author} {\bibfnamefont {Y.}~\bibnamefont {Zhou}}, \bibinfo
  {author} {\bibfnamefont {H.-X.}\ \bibnamefont {Chen}}, \bibinfo {author}
  {\bibfnamefont {J.-J.}\ \bibnamefont {Xie}},\ and\ \bibinfo {author}
  {\bibfnamefont {L.-S.}\ \bibnamefont {Geng}},\ }\bibfield  {title} {\bibinfo
  {title} {{Dynamically generated $J^P=1/2^-(3/2^-)$ singly charmed and bottom
  heavy baryons}},\ }\href {https://doi.org/10.1103/PhysRevD.92.014036}
  {\bibfield  {journal} {\bibinfo  {journal} {Phys. Rev. D}\ }\textbf {\bibinfo
  {volume} {92}},\ \bibinfo {pages} {014036} (\bibinfo {year} {2015})},\
  \Eprint {https://arxiv.org/abs/1409.3133} {arXiv:1409.3133 [hep-ph]}
  \BibitemShut {NoStop}%
\bibitem [{\citenamefont {Lu}\ \emph {et~al.}(2016)\citenamefont {Lu},
  \citenamefont {Chen}, \citenamefont {Guo}, \citenamefont {Nieves},
  \citenamefont {Xie},\ and\ \citenamefont {Geng}}]{Lu:2016gev}%
  \BibitemOpen
  \bibfield  {author} {\bibinfo {author} {\bibfnamefont {J.-X.}\ \bibnamefont
  {Lu}}, \bibinfo {author} {\bibfnamefont {H.-X.}\ \bibnamefont {Chen}},
  \bibinfo {author} {\bibfnamefont {Z.-H.}\ \bibnamefont {Guo}}, \bibinfo
  {author} {\bibfnamefont {J.}~\bibnamefont {Nieves}}, \bibinfo {author}
  {\bibfnamefont {J.-J.}\ \bibnamefont {Xie}},\ and\ \bibinfo {author}
  {\bibfnamefont {L.-S.}\ \bibnamefont {Geng}},\ }\bibfield  {title} {\bibinfo
  {title} {{$\Lambda_c(2595)$ resonance as a dynamically generated state: The
  compositeness condition and the large $N_c$ evolution}},\ }\href
  {https://doi.org/10.1103/PhysRevD.93.114028} {\bibfield  {journal} {\bibinfo
  {journal} {Phys. Rev. D}\ }\textbf {\bibinfo {volume} {93}},\ \bibinfo
  {pages} {114028} (\bibinfo {year} {2016})},\ \Eprint
  {https://arxiv.org/abs/1603.05388} {arXiv:1603.05388 [hep-ph]} \BibitemShut
  {NoStop}%
\bibitem [{\citenamefont {Oller}\ and\ \citenamefont
  {Meissner}(2001)}]{Oller:2000fj}%
  \BibitemOpen
  \bibfield  {author} {\bibinfo {author} {\bibfnamefont {J.~A.}\ \bibnamefont
  {Oller}}\ and\ \bibinfo {author} {\bibfnamefont {U.~G.}\ \bibnamefont
  {Meissner}},\ }\bibfield  {title} {\bibinfo {title} {{Chiral dynamics in the
  presence of bound states: Kaon nucleon interactions revisited}},\ }\href
  {https://doi.org/10.1016/S0370-2693(01)00078-8} {\bibfield  {journal}
  {\bibinfo  {journal} {Phys. Lett. B}\ }\textbf {\bibinfo {volume} {500}},\
  \bibinfo {pages} {263} (\bibinfo {year} {2001})},\ \Eprint
  {https://arxiv.org/abs/hep-ph/0011146} {arXiv:hep-ph/0011146} \BibitemShut
  {NoStop}%
\bibitem [{\citenamefont {Garcia-Recio}\ \emph {et~al.}(2003)\citenamefont
  {Garcia-Recio}, \citenamefont {Nieves}, \citenamefont {Ruiz~Arriola},\ and\
  \citenamefont {Vicente~Vacas}}]{Garcia-Recio:2002yxy}%
  \BibitemOpen
  \bibfield  {author} {\bibinfo {author} {\bibfnamefont {C.}~\bibnamefont
  {Garcia-Recio}}, \bibinfo {author} {\bibfnamefont {J.}~\bibnamefont
  {Nieves}}, \bibinfo {author} {\bibfnamefont {E.}~\bibnamefont
  {Ruiz~Arriola}},\ and\ \bibinfo {author} {\bibfnamefont {M.~J.}\ \bibnamefont
  {Vicente~Vacas}},\ }\bibfield  {title} {\bibinfo {title} {{S = -1 meson
  baryon unitarized coupled channel chiral perturbation theory and the S(01)
  Lambda(1405) and Lambda(1670) resonances}},\ }\href
  {https://doi.org/10.1103/PhysRevD.67.076009} {\bibfield  {journal} {\bibinfo
  {journal} {Phys. Rev. D}\ }\textbf {\bibinfo {volume} {67}},\ \bibinfo
  {pages} {076009} (\bibinfo {year} {2003})},\ \Eprint
  {https://arxiv.org/abs/hep-ph/0210311} {arXiv:hep-ph/0210311} \BibitemShut
  {NoStop}%
\bibitem [{\citenamefont {Garcia-Recio}\ \emph {et~al.}(2004)\citenamefont
  {Garcia-Recio}, \citenamefont {Lutz},\ and\ \citenamefont
  {Nieves}}]{Garcia-Recio:2003ejq}%
  \BibitemOpen
  \bibfield  {author} {\bibinfo {author} {\bibfnamefont {C.}~\bibnamefont
  {Garcia-Recio}}, \bibinfo {author} {\bibfnamefont {M.~F.~M.}\ \bibnamefont
  {Lutz}},\ and\ \bibinfo {author} {\bibfnamefont {J.}~\bibnamefont {Nieves}},\
  }\bibfield  {title} {\bibinfo {title} {{Quark mass dependence of s wave
  baryon resonances}},\ }\href {https://doi.org/10.1016/j.physletb.2003.11.073}
  {\bibfield  {journal} {\bibinfo  {journal} {Phys. Lett. B}\ }\textbf
  {\bibinfo {volume} {582}},\ \bibinfo {pages} {49} (\bibinfo {year} {2004})},\
  \Eprint {https://arxiv.org/abs/nucl-th/0305100} {arXiv:nucl-th/0305100}
  \BibitemShut {NoStop}%
\bibitem [{\citenamefont {Hyodo}\ and\ \citenamefont
  {Jido}(2012)}]{Hyodo:2011ur}%
  \BibitemOpen
  \bibfield  {author} {\bibinfo {author} {\bibfnamefont {T.}~\bibnamefont
  {Hyodo}}\ and\ \bibinfo {author} {\bibfnamefont {D.}~\bibnamefont {Jido}},\
  }\bibfield  {title} {\bibinfo {title} {{The nature of the Lambda(1405)
  resonance in chiral dynamics}},\ }\href
  {https://doi.org/10.1016/j.ppnp.2011.07.002} {\bibfield  {journal} {\bibinfo
  {journal} {Prog. Part. Nucl. Phys.}\ }\textbf {\bibinfo {volume} {67}},\
  \bibinfo {pages} {55} (\bibinfo {year} {2012})},\ \Eprint
  {https://arxiv.org/abs/1104.4474} {arXiv:1104.4474 [nucl-th]} \BibitemShut
  {NoStop}%
\bibitem [{\citenamefont {Nieves}\ \emph {et~al.}(2019)\citenamefont {Nieves},
  \citenamefont {Pavao},\ and\ \citenamefont {Sakai}}]{Nieves:2019kdh}%
  \BibitemOpen
  \bibfield  {author} {\bibinfo {author} {\bibfnamefont {J.}~\bibnamefont
  {Nieves}}, \bibinfo {author} {\bibfnamefont {R.}~\bibnamefont {Pavao}},\ and\
  \bibinfo {author} {\bibfnamefont {S.}~\bibnamefont {Sakai}},\ }\bibfield
  {title} {\bibinfo {title} {{$\Lambda _b$ decays into $\Lambda _c^*\ell
  \bar{\nu }_\ell $ and $\Lambda _c^*\pi ^-$ $[\Lambda _c^*=\Lambda _c(2595)$
  and $\Lambda _c(2625)]$ and heavy quark spin symmetry}},\ }\href
  {https://doi.org/10.1140/epjc/s10052-019-6929-7} {\bibfield  {journal}
  {\bibinfo  {journal} {Eur. Phys. J. C}\ }\textbf {\bibinfo {volume} {79}},\
  \bibinfo {pages} {417} (\bibinfo {year} {2019})},\ \Eprint
  {https://arxiv.org/abs/1903.11911} {arXiv:1903.11911 [hep-ph]} \BibitemShut
  {NoStop}%
\bibitem [{\citenamefont {Liang}\ \emph {et~al.}(2015)\citenamefont {Liang},
  \citenamefont {Uchino}, \citenamefont {Xiao},\ and\ \citenamefont
  {Oset}}]{Liang:2014kra}%
  \BibitemOpen
  \bibfield  {author} {\bibinfo {author} {\bibfnamefont {W.~H.}\ \bibnamefont
  {Liang}}, \bibinfo {author} {\bibfnamefont {T.}~\bibnamefont {Uchino}},
  \bibinfo {author} {\bibfnamefont {C.~W.}\ \bibnamefont {Xiao}},\ and\
  \bibinfo {author} {\bibfnamefont {E.}~\bibnamefont {Oset}},\ }\bibfield
  {title} {\bibinfo {title} {{Baryon states with open charm in the extended
  local hidden gauge approach}},\ }\href
  {https://doi.org/10.1140/epja/i2015-15016-1} {\bibfield  {journal} {\bibinfo
  {journal} {Eur. Phys. J. A}\ }\textbf {\bibinfo {volume} {51}},\ \bibinfo
  {pages} {16} (\bibinfo {year} {2015})},\ \Eprint
  {https://arxiv.org/abs/1402.5293} {arXiv:1402.5293 [hep-ph]} \BibitemShut
  {NoStop}%
\bibitem [{\citenamefont {Nieves}\ and\ \citenamefont
  {Pavao}(2020)}]{Nieves:2019nol}%
  \BibitemOpen
  \bibfield  {author} {\bibinfo {author} {\bibfnamefont {J.}~\bibnamefont
  {Nieves}}\ and\ \bibinfo {author} {\bibfnamefont {R.}~\bibnamefont {Pavao}},\
  }\bibfield  {title} {\bibinfo {title} {{Nature of the lowest-lying odd parity
  charmed baryon $\Lambda_c(2595)$ and $\Lambda_c(2625)$ resonances}},\ }\href
  {https://doi.org/10.1103/PhysRevD.101.014018} {\bibfield  {journal} {\bibinfo
   {journal} {Phys. Rev. D}\ }\textbf {\bibinfo {volume} {101}},\ \bibinfo
  {pages} {014018} (\bibinfo {year} {2020})},\ \Eprint
  {https://arxiv.org/abs/1907.05747} {arXiv:1907.05747 [hep-ph]} \BibitemShut
  {NoStop}%
\bibitem [{\citenamefont {Mannel}\ \emph {et~al.}(1991)\citenamefont {Mannel},
  \citenamefont {Roberts},\ and\ \citenamefont {Ryzak}}]{Mannel:1990vg}%
  \BibitemOpen
  \bibfield  {author} {\bibinfo {author} {\bibfnamefont {T.}~\bibnamefont
  {Mannel}}, \bibinfo {author} {\bibfnamefont {W.}~\bibnamefont {Roberts}},\
  and\ \bibinfo {author} {\bibfnamefont {Z.}~\bibnamefont {Ryzak}},\ }\bibfield
   {title} {\bibinfo {title} {{Baryons in the heavy quark effective theory}},\
  }\href {https://doi.org/10.1016/0550-3213(91)90301-D} {\bibfield  {journal}
  {\bibinfo  {journal} {Nucl. Phys. B}\ }\textbf {\bibinfo {volume} {355}},\
  \bibinfo {pages} {38} (\bibinfo {year} {1991})}\BibitemShut {NoStop}%
\bibitem [{\citenamefont {Isgur}\ and\ \citenamefont
  {Wise}(1991)}]{Isgur:1990pm}%
  \BibitemOpen
  \bibfield  {author} {\bibinfo {author} {\bibfnamefont {N.}~\bibnamefont
  {Isgur}}\ and\ \bibinfo {author} {\bibfnamefont {M.~B.}\ \bibnamefont
  {Wise}},\ }\bibfield  {title} {\bibinfo {title} {{Heavy baryon weak
  form-factors}},\ }\href {https://doi.org/10.1016/0550-3213(91)90518-3}
  {\bibfield  {journal} {\bibinfo  {journal} {Nucl. Phys. B}\ }\textbf
  {\bibinfo {volume} {348}},\ \bibinfo {pages} {276} (\bibinfo {year}
  {1991})}\BibitemShut {NoStop}%
\bibitem [{\citenamefont {Roberts}(1993)}]{Roberts:1992xm}%
  \BibitemOpen
  \bibfield  {author} {\bibinfo {author} {\bibfnamefont {W.}~\bibnamefont
  {Roberts}},\ }\bibfield  {title} {\bibinfo {title} {{Semileptonic decays of
  heavy Lambda's into excited baryons}},\ }\href
  {https://doi.org/10.1016/0550-3213(93)90331-I} {\bibfield  {journal}
  {\bibinfo  {journal} {Nucl. Phys. B}\ }\textbf {\bibinfo {volume} {389}},\
  \bibinfo {pages} {549} (\bibinfo {year} {1993})}\BibitemShut {NoStop}%
\bibitem [{\citenamefont {Meinel}\ and\ \citenamefont
  {Rendon}(2022)}]{Meinel:2021mdj}%
  \BibitemOpen
  \bibfield  {author} {\bibinfo {author} {\bibfnamefont {S.}~\bibnamefont
  {Meinel}}\ and\ \bibinfo {author} {\bibfnamefont {G.}~\bibnamefont
  {Rendon}},\ }\bibfield  {title} {\bibinfo {title}
  {{\ensuremath{\Lambda}c\textrightarrow{}\ensuremath{\Lambda}*(1520) form
  factors from lattice QCD and improved analysis of the
  \ensuremath{\Lambda}b\textrightarrow{}\ensuremath{\Lambda}*(1520) and
  \ensuremath{\Lambda}b\textrightarrow{}\ensuremath{\Lambda}c*(2595,2625) form
  factors}},\ }\href {https://doi.org/10.1103/PhysRevD.105.054511} {\bibfield
  {journal} {\bibinfo  {journal} {Phys. Rev. D}\ }\textbf {\bibinfo {volume}
  {105}},\ \bibinfo {pages} {054511} (\bibinfo {year} {2022})},\ \Eprint
  {https://arxiv.org/abs/2107.13140} {arXiv:2107.13140 [hep-lat]} \BibitemShut
  {NoStop}%
\bibitem [{\citenamefont {Gutsche}\ \emph {et~al.}(2017)\citenamefont
  {Gutsche}, \citenamefont {Ivanov}, \citenamefont {K\"orner}, \citenamefont
  {Lyubovitskij}, \citenamefont {Lyubushkin},\ and\ \citenamefont
  {Santorelli}}]{Gutsche:2017wag}%
  \BibitemOpen
  \bibfield  {author} {\bibinfo {author} {\bibfnamefont {T.}~\bibnamefont
  {Gutsche}}, \bibinfo {author} {\bibfnamefont {M.~A.}\ \bibnamefont {Ivanov}},
  \bibinfo {author} {\bibfnamefont {J.~G.}\ \bibnamefont {K\"orner}}, \bibinfo
  {author} {\bibfnamefont {V.~E.}\ \bibnamefont {Lyubovitskij}}, \bibinfo
  {author} {\bibfnamefont {V.~V.}\ \bibnamefont {Lyubushkin}},\ and\ \bibinfo
  {author} {\bibfnamefont {P.}~\bibnamefont {Santorelli}},\ }\bibfield  {title}
  {\bibinfo {title} {{Theoretical description of the decays $\Lambda_b \to
  \Lambda^{(\ast)}(\frac12^\pm,\frac32^\pm) + J/\psi$}},\ }\href
  {https://doi.org/10.1103/PhysRevD.96.013003} {\bibfield  {journal} {\bibinfo
  {journal} {Phys. Rev. D}\ }\textbf {\bibinfo {volume} {96}},\ \bibinfo
  {pages} {013003} (\bibinfo {year} {2017})},\ \Eprint
  {https://arxiv.org/abs/1705.07299} {arXiv:1705.07299 [hep-ph]} \BibitemShut
  {NoStop}%
\bibitem [{\citenamefont {Be\v{c}irevi\'c}\ \emph {et~al.}(2020)\citenamefont
  {Be\v{c}irevi\'c}, \citenamefont {Le~Yaouanc}, \citenamefont {Mor\'enas},\
  and\ \citenamefont {Oliver}}]{Becirevic:2020nmb}%
  \BibitemOpen
  \bibfield  {author} {\bibinfo {author} {\bibfnamefont {D.}~\bibnamefont
  {Be\v{c}irevi\'c}}, \bibinfo {author} {\bibfnamefont {A.}~\bibnamefont
  {Le~Yaouanc}}, \bibinfo {author} {\bibfnamefont {V.}~\bibnamefont
  {Mor\'enas}},\ and\ \bibinfo {author} {\bibfnamefont {L.}~\bibnamefont
  {Oliver}},\ }\bibfield  {title} {\bibinfo {title} {{Heavy baryon wave
  functions, Bakamjian-Thomas approach to form factors, and observables in
  ${\Lambda_b \to \Lambda_c\left({1 \over 2}^\pm \right) \ell \overline{\nu}}$
  transitions}},\ }\href {https://doi.org/10.1103/PhysRevD.102.094023}
  {\bibfield  {journal} {\bibinfo  {journal} {Phys. Rev. D}\ }\textbf {\bibinfo
  {volume} {102}},\ \bibinfo {pages} {094023} (\bibinfo {year} {2020})},\
  \Eprint {https://arxiv.org/abs/2006.07130} {arXiv:2006.07130 [hep-ph]}
  \BibitemShut {NoStop}%
\bibitem [{\citenamefont {Bowler}\ \emph {et~al.}(1998)\citenamefont {Bowler},
  \citenamefont {Kenway}, \citenamefont {Lellouch}, \citenamefont {Nieves},
  \citenamefont {Oliveira}, \citenamefont {Richards}, \citenamefont
  {Sachrajda}, \citenamefont {Stella},\ and\ \citenamefont
  {Ueberholz}}]{Bowler:1997ej}%
  \BibitemOpen
  \bibfield  {author} {\bibinfo {author} {\bibfnamefont {K.~C.}\ \bibnamefont
  {Bowler}}, \bibinfo {author} {\bibfnamefont {R.~D.}\ \bibnamefont {Kenway}},
  \bibinfo {author} {\bibfnamefont {L.}~\bibnamefont {Lellouch}}, \bibinfo
  {author} {\bibfnamefont {J.}~\bibnamefont {Nieves}}, \bibinfo {author}
  {\bibfnamefont {O.}~\bibnamefont {Oliveira}}, \bibinfo {author}
  {\bibfnamefont {D.~G.}\ \bibnamefont {Richards}}, \bibinfo {author}
  {\bibfnamefont {C.~T.}\ \bibnamefont {Sachrajda}}, \bibinfo {author}
  {\bibfnamefont {N.}~\bibnamefont {Stella}},\ and\ \bibinfo {author}
  {\bibfnamefont {P.}~\bibnamefont {Ueberholz}} (\bibinfo {collaboration}
  {UKQCD}),\ }\bibfield  {title} {\bibinfo {title} {{First lattice study of
  semileptonic decays of Lambda-b and Xi-b baryons}},\ }\href
  {https://doi.org/10.1103/PhysRevD.57.6948} {\bibfield  {journal} {\bibinfo
  {journal} {Phys. Rev. D}\ }\textbf {\bibinfo {volume} {57}},\ \bibinfo
  {pages} {6948} (\bibinfo {year} {1998})},\ \Eprint
  {https://arxiv.org/abs/hep-lat/9709028} {arXiv:hep-lat/9709028} \BibitemShut
  {NoStop}%
\bibitem [{\citenamefont {Gottlieb}\ and\ \citenamefont
  {Tamhankar}(2003)}]{Gottlieb:2003yb}%
  \BibitemOpen
  \bibfield  {author} {\bibinfo {author} {\bibfnamefont {S.~A.}\ \bibnamefont
  {Gottlieb}}\ and\ \bibinfo {author} {\bibfnamefont {S.}~\bibnamefont
  {Tamhankar}},\ }\bibfield  {title} {\bibinfo {title} {{A Lattice study of
  Lambda(b) semileptonic decay}},\ }\href
  {https://doi.org/10.1016/S0920-5632(03)01612-8} {\bibfield  {journal}
  {\bibinfo  {journal} {Nucl. Phys. B Proc. Suppl.}\ }\textbf {\bibinfo
  {volume} {119}},\ \bibinfo {pages} {644} (\bibinfo {year} {2003})},\ \Eprint
  {https://arxiv.org/abs/hep-lat/0301022} {arXiv:hep-lat/0301022} \BibitemShut
  {NoStop}%
\bibitem [{\citenamefont {Detmold}\ \emph {et~al.}(2015)\citenamefont
  {Detmold}, \citenamefont {Lehner},\ and\ \citenamefont
  {Meinel}}]{Detmold:2015aaa}%
  \BibitemOpen
  \bibfield  {author} {\bibinfo {author} {\bibfnamefont {W.}~\bibnamefont
  {Detmold}}, \bibinfo {author} {\bibfnamefont {C.}~\bibnamefont {Lehner}},\
  and\ \bibinfo {author} {\bibfnamefont {S.}~\bibnamefont {Meinel}},\
  }\bibfield  {title} {\bibinfo {title} {{$\Lambda_b \to p \ell^-
  \bar{\nu}_\ell$ and $\Lambda_b \to \Lambda_c \ell^- \bar{\nu}_\ell$ form
  factors from lattice QCD with relativistic heavy quarks}},\ }\href
  {https://doi.org/10.1103/PhysRevD.92.034503} {\bibfield  {journal} {\bibinfo
  {journal} {Phys. Rev. D}\ }\textbf {\bibinfo {volume} {92}},\ \bibinfo
  {pages} {034503} (\bibinfo {year} {2015})},\ \Eprint
  {https://arxiv.org/abs/1503.01421} {arXiv:1503.01421 [hep-lat]} \BibitemShut
  {NoStop}%
\bibitem [{\citenamefont {Datta}\ \emph {et~al.}(2017)\citenamefont {Datta},
  \citenamefont {Kamali}, \citenamefont {Meinel},\ and\ \citenamefont
  {Rashed}}]{Datta:2017aue}%
  \BibitemOpen
  \bibfield  {author} {\bibinfo {author} {\bibfnamefont {A.}~\bibnamefont
  {Datta}}, \bibinfo {author} {\bibfnamefont {S.}~\bibnamefont {Kamali}},
  \bibinfo {author} {\bibfnamefont {S.}~\bibnamefont {Meinel}},\ and\ \bibinfo
  {author} {\bibfnamefont {A.}~\bibnamefont {Rashed}},\ }\bibfield  {title}
  {\bibinfo {title} {{Phenomenology of $ {\Lambda}_b\to {\Lambda}_c\tau
  {\overline{\nu}}_{\tau } $ using lattice QCD calculations}},\ }\href
  {https://doi.org/10.1007/JHEP08(2017)131} {\bibfield  {journal} {\bibinfo
  {journal} {JHEP}\ }\textbf {\bibinfo {volume} {08}},\ \bibinfo {pages}
  {131}},\ \Eprint {https://arxiv.org/abs/1702.02243} {arXiv:1702.02243
  [hep-ph]} \BibitemShut {NoStop}%
\bibitem [{\citenamefont {Aaij}\ \emph {et~al.}(2017)\citenamefont {Aaij} \emph
  {et~al.}}]{LHCb:2017vhq}%
  \BibitemOpen
  \bibfield  {author} {\bibinfo {author} {\bibfnamefont {R.}~\bibnamefont
  {Aaij}} \emph {et~al.} (\bibinfo {collaboration} {LHCb}),\ }\bibfield
  {title} {\bibinfo {title} {{Measurement of the shape of the
  $\Lambda_b^0\to\Lambda_c^+ \mu^- \overline{\nu}_{\mu}$ differential decay
  rate}},\ }\href {https://doi.org/10.1103/PhysRevD.96.112005} {\bibfield
  {journal} {\bibinfo  {journal} {Phys. Rev. D}\ }\textbf {\bibinfo {volume}
  {96}},\ \bibinfo {pages} {112005} (\bibinfo {year} {2017})},\ \Eprint
  {https://arxiv.org/abs/1709.01920} {arXiv:1709.01920 [hep-ex]} \BibitemShut
  {NoStop}%
\bibitem [{\citenamefont {Shifman}\ \emph {et~al.}(1995)\citenamefont
  {Shifman}, \citenamefont {Uraltsev},\ and\ \citenamefont
  {Vainshtein}}]{Shifman:1994jh}%
  \BibitemOpen
  \bibfield  {author} {\bibinfo {author} {\bibfnamefont {M.~A.}\ \bibnamefont
  {Shifman}}, \bibinfo {author} {\bibfnamefont {N.~G.}\ \bibnamefont
  {Uraltsev}},\ and\ \bibinfo {author} {\bibfnamefont {A.~I.}\ \bibnamefont
  {Vainshtein}},\ }\bibfield  {title} {\bibinfo {title} {{V(cb) from OPE sum
  rules for heavy flavor transitions}},\ }\href
  {https://doi.org/10.1103/PhysRevD.52.3149} {\bibfield  {journal} {\bibinfo
  {journal} {Phys. Rev. D}\ }\textbf {\bibinfo {volume} {51}},\ \bibinfo
  {pages} {2217} (\bibinfo {year} {1995})},\ \bibinfo {note} {[Erratum:
  Phys.Rev.D 52, 3149 (1995)]},\ \Eprint {https://arxiv.org/abs/hep-ph/9405207}
  {arXiv:hep-ph/9405207} \BibitemShut {NoStop}%
\bibitem [{\citenamefont {Bigi}\ \emph {et~al.}(1995)\citenamefont {Bigi},
  \citenamefont {Shifman}, \citenamefont {Uraltsev},\ and\ \citenamefont
  {Vainshtein}}]{Bigi:1994ga}%
  \BibitemOpen
  \bibfield  {author} {\bibinfo {author} {\bibfnamefont {I.~I.~Y.}\
  \bibnamefont {Bigi}}, \bibinfo {author} {\bibfnamefont {M.~A.}\ \bibnamefont
  {Shifman}}, \bibinfo {author} {\bibfnamefont {N.~G.}\ \bibnamefont
  {Uraltsev}},\ and\ \bibinfo {author} {\bibfnamefont {A.~I.}\ \bibnamefont
  {Vainshtein}},\ }\bibfield  {title} {\bibinfo {title} {{Sum rules for heavy
  flavor transitions in the SV limit}},\ }\href
  {https://doi.org/10.1103/PhysRevD.52.196} {\bibfield  {journal} {\bibinfo
  {journal} {Phys. Rev. D}\ }\textbf {\bibinfo {volume} {52}},\ \bibinfo
  {pages} {196} (\bibinfo {year} {1995})},\ \Eprint
  {https://arxiv.org/abs/hep-ph/9405410} {arXiv:hep-ph/9405410} \BibitemShut
  {NoStop}%
\bibitem [{\citenamefont {Falk}(1992)}]{Falk:1991nq}%
  \BibitemOpen
  \bibfield  {author} {\bibinfo {author} {\bibfnamefont {A.~F.}\ \bibnamefont
  {Falk}},\ }\bibfield  {title} {\bibinfo {title} {{Hadrons of arbitrary spin
  in the heavy quark effective theory}},\ }\href
  {https://doi.org/10.1016/0550-3213(92)90004-U} {\bibfield  {journal}
  {\bibinfo  {journal} {Nucl. Phys. B}\ }\textbf {\bibinfo {volume} {378}},\
  \bibinfo {pages} {79} (\bibinfo {year} {1992})}\BibitemShut {NoStop}%
\bibitem [{\citenamefont {Neubert}(1994)}]{Neubert:1993mb}%
  \BibitemOpen
  \bibfield  {author} {\bibinfo {author} {\bibfnamefont {M.}~\bibnamefont
  {Neubert}},\ }\bibfield  {title} {\bibinfo {title} {{Heavy quark symmetry}},\
  }\href {https://doi.org/10.1016/0370-1573(94)90091-4} {\bibfield  {journal}
  {\bibinfo  {journal} {Phys. Rept.}\ }\textbf {\bibinfo {volume} {245}},\
  \bibinfo {pages} {259} (\bibinfo {year} {1994})},\ \Eprint
  {https://arxiv.org/abs/hep-ph/9306320} {arXiv:hep-ph/9306320} \BibitemShut
  {NoStop}%
\bibitem [{\citenamefont {Aoki}\ \emph {et~al.}(2011)\citenamefont {Aoki} \emph
  {et~al.}}]{RBC:2010qam}%
  \BibitemOpen
  \bibfield  {author} {\bibinfo {author} {\bibfnamefont {Y.}~\bibnamefont
  {Aoki}} \emph {et~al.} (\bibinfo {collaboration} {RBC, UKQCD}),\ }\bibfield
  {title} {\bibinfo {title} {{Continuum Limit Physics from 2+1 Flavor Domain
  Wall QCD}},\ }\href {https://doi.org/10.1103/PhysRevD.83.074508} {\bibfield
  {journal} {\bibinfo  {journal} {Phys. Rev. D}\ }\textbf {\bibinfo {volume}
  {83}},\ \bibinfo {pages} {074508} (\bibinfo {year} {2011})},\ \Eprint
  {https://arxiv.org/abs/1011.0892} {arXiv:1011.0892 [hep-lat]} \BibitemShut
  {NoStop}%
\bibitem [{\citenamefont {Blum}\ \emph {et~al.}(2016)\citenamefont {Blum} \emph
  {et~al.}}]{RBC:2014ntl}%
  \BibitemOpen
  \bibfield  {author} {\bibinfo {author} {\bibfnamefont {T.}~\bibnamefont
  {Blum}} \emph {et~al.} (\bibinfo {collaboration} {RBC, UKQCD}),\ }\bibfield
  {title} {\bibinfo {title} {{Domain wall QCD with physical quark masses}},\
  }\href {https://doi.org/10.1103/PhysRevD.93.074505} {\bibfield  {journal}
  {\bibinfo  {journal} {Phys. Rev. D}\ }\textbf {\bibinfo {volume} {93}},\
  \bibinfo {pages} {074505} (\bibinfo {year} {2016})},\ \Eprint
  {https://arxiv.org/abs/1411.7017} {arXiv:1411.7017 [hep-lat]} \BibitemShut
  {NoStop}%
\bibitem [{\citenamefont {Zyla}\ \emph {et~al.}(2020)\citenamefont {Zyla} \emph
  {et~al.}}]{ParticleDataGroup:2020ssz}%
  \BibitemOpen
  \bibfield  {author} {\bibinfo {author} {\bibfnamefont {P.~A.}\ \bibnamefont
  {Zyla}} \emph {et~al.} (\bibinfo {collaboration} {Particle Data Group}),\
  }\bibfield  {title} {\bibinfo {title} {{Review of Particle Physics}},\ }\href
  {https://doi.org/10.1093/ptep/ptaa104} {\bibfield  {journal} {\bibinfo
  {journal} {PTEP}\ }\textbf {\bibinfo {volume} {2020}},\ \bibinfo {pages}
  {083C01} (\bibinfo {year} {2020})}\BibitemShut {NoStop}%
\bibitem [{\citenamefont {Aaltonen}\ \emph {et~al.}(2009)\citenamefont
  {Aaltonen} \emph {et~al.}}]{CDF:2008hqh}%
  \BibitemOpen
  \bibfield  {author} {\bibinfo {author} {\bibfnamefont {T.}~\bibnamefont
  {Aaltonen}} \emph {et~al.} (\bibinfo {collaboration} {CDF}),\ }\bibfield
  {title} {\bibinfo {title} {{First Measurement of the Ratio of Branching
  Fractions $B(\Lambda^0_b \to \Lambda^+_{c} \mu^{-} \bar{\nu}_\mu /
  B(Lambda^0_b \to \Lambda^+_{c} \pi^{-})$}},\ }\href
  {https://doi.org/10.1103/PhysRevD.79.032001} {\bibfield  {journal} {\bibinfo
  {journal} {Phys. Rev. D}\ }\textbf {\bibinfo {volume} {79}},\ \bibinfo
  {pages} {032001} (\bibinfo {year} {2009})},\ \Eprint
  {https://arxiv.org/abs/0810.3213} {arXiv:0810.3213 [hep-ex]} \BibitemShut
  {NoStop}%
\end{thebibliography}%

\end{document}